\documentclass[acmsmall]{acmart}
\usepackage{mathtools}
\usepackage{amsmath}
\usepackage{algorithm}
\usepackage[noend]{algpseudocode} 
\usepackage{pdfpages}
\usepackage{float}
\usepackage{listings}
\usepackage{xpatch}
\usepackage{forest}
\usepackage{comment} 
\usepackage{textcomp}
\usepackage{xspace}
\usetikzlibrary{positioning, arrows.meta, shadows}
\usepackage{nicematrix}
\usepackage{wrapfig}
\usepackage{pifont}

\newcommand{\kleeneStar}{^\texttt{*}}
\newcommand{\kleenePlus}{^\texttt{+}}
\newcommand{\tensor}{t}

\newcommand{\const}{const}
\newcommand{\template}{\tau}
\newcommand{\templateset}{\mathcal{T}}

\newcommand{\sys}{\textsc{STAGG}\xspace}
\newcommand{\systd}{\textsc{STAGG}$^{TD}$\xspace}
\newcommand{\sysbu}{\textsc{STAGG}$^{BU}$\xspace}
\newcommand{\astar}{A$^*$ }
\newcommand{\nt}{V}
\newcommand{\T}{\Sigma}
\newcommand{\rules}{R}
\newcommand{\start}{S}
\newcommand{\weight}{W}
\newcommand{\grammar}{G}
\newcommand{\prob}{\mathbb{P}}
\newcommand{\Penalty}{\mathcal{X}}
\newcommand{\languageOfGrammar}{\ensuremath{\mathcal{L}}}
\newcommand{\tikzxmark}{%
\tikz[scale=0.23] {
    \draw[line width=0.7,line cap=round] (0,0) to [bend left=6] (1,1);
    \draw[line width=0.7,line cap=round] (0.2,0.95) to [bend right=3] (0.8,0.05);
}}
\newcommand{\tikzcmark}{%
\tikz[scale=0.23] {
    \draw[line width=0.7,line cap=round] (0.25,0) to [bend left=10] (1,1);
    \draw[line width=0.8,line cap=round] (0,0.35) to [bend right=1] (0.23,0);
}}
\newcommand{\src}{p_s}
\newcommand{\target}{p_t}
\newcommand{\inputs}{\vec{x}}

\newcounter{prompt}
\renewcommand{\theprompt}{\arabic{prompt}}

\newcounter{response}
\renewcommand{\theresponse}{\arabic{response}}

\newcounter{program}

\newcounter{eg}
\renewcommand{\theeg}{\arabic{eg}}

\newcounter{nobnf}
\renewcommand{\thenobnf}{\arabic{nobnf}}
\newcounter{bnf}
\renewcommand{\thebnf}{\arabic{bnf}}

\lstnewenvironment{prompt}[1][]{%
  \let\originalthelstlisting\thelstlisting%
  \renewcommand{\thelstlisting}{\theprompt}%
  \refstepcounter{prompt}%
  \lstset{language=text, captionpos=b, #1}%
}{%
  \let\thelstlisting\originalthelstlisting%
}

\lstnewenvironment{response}[1][]{%
  \let\originalthelstlisting\thelstlisting%
  \renewcommand{\thelstlisting}{\theresponse}%
  \refstepcounter{response}%
  \lstset{language=text, captionpos=b, #1}%
}{%
  \let\thelstlisting\originalthelstlisting%
}

\newcommand{\lstlistprogram}{List of Program}
\newcommand{\lstlistofprogram}{\bgroup
    \let\contentsname\lstlistprogram
    \let\lst@temp\@starttoc \def\@starttoc##1{\lst@temp{lop}}%
    \tableofcontents \egroup}
\lstnewenvironment{program}[1][]{%
  \xpatchcmd*{\lst@MakeCaption}{lol}{lop}{}{}%
  \lstset{language=text, captionpos=b, #1}}
  {}

\lstnewenvironment{eg}[1][]{%
  \let\originalthelstlisting\thelstlisting%
  \renewcommand{\thelstlisting}{\theeg}%
  \refstepcounter{eg}%
  \lstset{language=myC, captionpos=b, #1}%
}{%
  \let\thelstlisting\originalthelstlisting%
}

\lstnewenvironment{bnf}[1][]{
  \let\originalthelstlisting\thelstlisting%
  \renewcommand{\thelstlisting}{\thebnf}%
  \refstepcounter{bnf}%
  \lstset{
    language=bnf, 
    captionpos=b, 
    mathescape=true, 
    basicstyle=\ttfamily\small, 
    #1
  }%
}{%
  \let\thelstlisting\originalthelstlisting%
}

\lstnewenvironment{nobnf}[1][]{
  \let\originalthelstlisting\thelstlisting%
  \renewcommand{\thelstlisting}{\thenobnf}%
  \refstepcounter{nobnf}%
  \lstset{
    language=bnf, 
    captionpos=b, 
    mathescape=true, 
    basicstyle=\ttfamily\small, 
    #1
  }%
}{%
  \let\thelstlisting\originalthelstlisting%
}
\definecolor{bluegray}{rgb}{0.0, 0.48, 0.65}
\lstdefinelanguage{text}{
    basicstyle=\ttfamily\small\color{bluegray},
    sensitive = false,
    keywords={},
    numbers=left,
    numberstyle=\ttfamily\scriptsize\color{Sepia},
    stepnumber=1,
    numbersep=8pt,
    showstringspaces=false,
    breaklines=true,  
    linewidth=0.95\linewidth,
    xleftmargin=0.05\linewidth,
}
\lstdefinelanguage{myC}{
    sensitive=true,
    keywords={function},
    keywords=[2]{Mat1, Mat2, Target, X, Y, N, t, m1, m2, p_m1, p_m2, p_t, i, f, k},
    keywords=[3]{void, int},
    keywords=[4]{for},
    keywordstyle=\ttfamily\bfseries\color{BrickRed},      
    keywordstyle=[2]\ttfamily\color{MidnightBlue},  
    keywordstyle=[3]\ttfamily\color{Plum},
    keywordstyle=[4]\ttfamily\color{Bittersweet},
    identifierstyle=\ttfamily\color{black},
    basicstyle=\ttfamily\small,
    numbers=left,
    numberstyle=\ttfamily\scriptsize\color{Sepia},
    stepnumber=1,
    numbersep=8pt,
    showstringspaces=false,
    breaklines=true,
    linewidth=0.95\linewidth,
    xleftmargin=0.05\linewidth,
    comment=[l]{//},                         
    morecomment=[s]{/*}{*/},                 
    commentstyle=\ttfamily\color{green},
    morestring=[b]",                         
    stringstyle=\ttfamily\color{black},
    literate=
      {0}{{\color{PineGreen}{\ttfamily 0}}}1
      {*}{{\normalsize{$*$}}}1
      {;}{{\color{black}{;}}}1
      {]}{{\color{black}{]}}}1
}

\lstdefinelanguage{response}{
    sensitive=true,
    keywords={function},
    keywords=[2]{},
    keywordstyle=\ttfamily\bfseries\color{red},      
    keywordstyle=[2]\ttfamily\color{Blue},           
    identifierstyle=\ttfamily\color{black},
    basicstyle=\ttfamily\small,
    numbers=left,
    numberstyle=\ttfamily\scriptsize\color{Sepia},
    stepnumber=1,
    numbersep=8pt,
    showstringspaces=false,
    breaklines=true,
    linewidth=0.95\linewidth,
    xleftmargin=0.05\linewidth,
    comment=[l]{//},                         
    morecomment=[s]{/*}{*/},                 
    commentstyle=\ttfamily\color{green},
    morestring=[b]",                         
    stringstyle=\ttfamily\color{brown},
    literate=
      {&}{{\color{Blue}{\ttfamily\&}}}1
      {0}{{\color{BlueViolet}{\ttfamily 0}}}1
      {*}{{\color{black}{\ttfamily $*$}}}1
      {+}{{\color{Blue}{\ttfamily +}}}1
      {-}{{\color{Blue}{\ttfamily -}}}1
      {/}{{\color{Blue}{\ttfamily /}}}1
      {<}{{\color{Blue}{\ttfamily <}}}1
      {>}{{\color{Blue}{\ttfamily >}}}1
      {=}{{\color{black}{\ttfamily =}}}1
      {==}{{\color{Blue}{\ttfamily ==}}}2
      {<=}{{\color{Blue}{\ttfamily <=}}}2
      {>=}{{\color{Blue}{\ttfamily >=}}}2
      {!=}{{\color{Blue}{\ttfamily !=}}}2
      {+=}{{\color{Blue}{\ttfamily +=}}}2
      {-=}{{\color{Blue}{\ttfamily -=}}}2
      {++}{{\color{Blue}{\ttfamily ++}}}2
      {--}{{\color{Blue}{\ttfamily --}}}2
      {[}{{\color{black}{\ttfamily [}}}1
      {]}{{\color{black}{\ttfamily ]}}}1
      {(}{{\color{black}{\ttfamily (}}}1
      {)}{{\color{black}{\ttfamily )}}}1
      {for}{{\color{Bittersweet}{\ttfamily for}}}3
      {int}{{\color{violet}{\ttfamily int}}}3
      {void}{{\color{violet}{\ttfamily void}}}4
}

\lstdefinelanguage{bnf}{
    basicstyle=\ttfamily\small,
    sensitive=false,
    morekeywords={PROGRAM, TENSOR, EXPR, INDEX-EXPR, INDEX-VAR, ID, CONSTANT, NUMBER, LETTER, DIGIT},
    keywordstyle=\color{Plum}\bfseries,
    morestring=[b]", 
    stringstyle=\color{Bittersweet}\ttfamily, 
    numbers=left,
    numberstyle=\ttfamily\scriptsize\color{Sepia},
    stepnumber=1,
    numbersep=8pt,
    showstringspaces=false,
    breaklines=true,
    linewidth=0.95\linewidth,
    xleftmargin=0.05\linewidth,
    mathescape=true, 
    moredelim=[is][\ttfamily\color{Plum}]{<}{>},
    aboveskip=0pt,
    belowskip=0pt,
    lineskip=1.5pt,
    literate=
        {EOL}{{\color{Bittersweet}{$\normalsize\varepsilon$}}}1
        {ADD}{{\color{Bittersweet}{"$+$"}}}3
        {SUB}{{\color{Bittersweet}{"$-$"}}}3
        {MUL}{{\color{Bittersweet}{"\normalsize{$*$}\small"}}}3
        {MULCons}{{\color{Bittersweet}{\normalsize{$*$ Const}\small}}}3
        {ADDCons}{{\color{Bittersweet}{\normalsize{$+$ Const}\small}}}3
        {SUBCons}{{\color{Bittersweet}{\normalsize{$-$ Const}\small}}}3
        {DIVCons}{{\color{Bittersweet}{\normalsize{$/$ Const}\small}}}3
        {DIV}{{\color{Bittersweet}{"$/$"}}}3
        {EQ}{{\color{Bittersweet}{"$=$"}}}3        
        {*}{{\color{black}{$^\texttt{*}$}}}1
        {+}{{\color{black}{$^+$}}}1
        {|}{{\color{black}{$\mid$}}}1
        {::=}{{\color{black}{\normalsize$\vcentcolon \vcentcolon=$}}}3 
        {[}{{\color{black}{[}}}1
        {(0}{{\color{black}{(0}}}2
        {(1}{{\color{black}{(1}}}2
}

\theoremstyle{definition}
\newtheorem{example}{Example}[section]
\AtBeginDocument{%
  }

\setcopyright{cc}
\setcctype{by}
\acmDOI{10.1145/3729330}
\acmYear{2025}
\acmJournal{PACMPL}
\acmVolume{9}
\acmNumber{PLDI}
\acmArticle{227}
\acmMonth{6}
\received{2024-11-15}
\received[accepted]{2025-03-06}
\begin{document}

\title{Guided Tensor Lifting}

\author{Yixuan Li}
\orcid{0009-0007-4619-3476}
\affiliation{%
  \institution{University of Edinburgh}
  \city{Edinburgh}
  \country{United Kingdom}
}
\email{yixuan.li.cs@ed.ac.uk}

\author{José Wesley de Souza Magalhães}
\orcid{0000-0003-2767-1130}
\affiliation{%
  \institution{University of Edinburgh}
  \city{Edinburgh}
  \country{United Kingdom}
}
\email{jwesley.magalhaes@ed.ac.uk}

\author{Alexander Brauckmann}
\orcid{0000-0001-5774-3970}
\affiliation{%
  \institution{University of Edinburgh}
  \city{Edinburgh}
  \country{United Kingdom}
}
\email{alexander.brauckmann@ed.ac.uk}

\author{Michael F. P. O'Boyle}
\orcid{0000-0003-1619-5052}
\affiliation{%
  \institution{University of Edinburgh}
  \city{Edinburgh}
  \country{United Kingdom}
}
\email{mob@inf.ed.ac.uk}

\author{Elizabeth Polgreen}
\orcid{0000-0001-9032-7661}
\affiliation{%
  \institution{University of Edinburgh}
  \city{Edinburgh}
  \country{United Kingdom}
}
\email{elizabeth.polgreen@ed.ac.uk}


\renewcommand{\shortauthors}{Y. Li, J. W. S. Magalh\~{a}es, A. Brauckmann, M. F. P. O'Boyle, E. Polgreen}


\begin{abstract}
Domain-specific languages (DSLs) for machine learning are revolutionizing the speed and efficiency of machine learning workloads as they enable users easy access to high-performance compiler optimizations and accelerators. However, to take advantage of these capabilities, a user must first translate their legacy code from the language it is currently written in, into the new DSL. The process of automatically lifting code into these DSLs has been identified by several recent works, which propose program synthesis as a solution. However, synthesis is expensive and struggles to scale without carefully designed and hard-wired heuristics. In this paper, we present an approach for lifting that combines an enumerative synthesis approach with a Large Language Model used to \emph{automatically} learn the domain-specific heuristics for program lifting, in the form of a probabilistic grammar. Our approach outperforms the state-of-the-art tools in this area, despite only using \emph{learned} heuristics.
\end{abstract}




\maketitle

\section{Introduction}

Recent years have witnessed rapid growth in the number and importance of machine learning workloads. While used in a diverse number of applications, their fundamental building block is tensor contractions, which dominate execution time.
For this reason,  a large number of specialized tensor domain-specific languages (DSLs) have appeared, capable of producing high-performance code (\cite{Kjolstad2017,harris2020array,paszke2019pytorch, abadi2016tensorflow, jax2018github,mlx2023}).
 Their associated compilers are capable of extracting domain-specific information to exploit hardware-specific features and vendor-tuned libraries.

To access such performance,  applications have to be written in one or more high-level DSLs. While this is acceptable for new applications, it means that
existing workloads written in standard programming languages are unable to directly access a platform's potential performance. While manually rewriting a program
to a DSL may be a worthwhile cost, it becomes a serious impediment if it has to be repeated for new emerging DSLs.
This problem of manual porting or {\em lifting}  existing code to higher-level DSLs has been identified by several recent works that propose automated techniques.
The most popular approaches use varying forms of program synthesis, where a
DSLs space is searched for a matching program ~(\cite{kamil2016verified, shi2022tf,c2tatco}).
 However, program synthesis
is expensive and struggles to scale to multi-dimensional tensor workloads.

To overcome this scalability issue, existing schemes rely on aggressive hard-wired heuristics that trade-off coverage for time. In C2TACO~\cite{c2tatco}, domain-specific polyhedral analysis is used to prune the search space. This works well on low-dimensional problems but suffers from exponential growth.
Similarly, in Tenspiler~\cite{qiu2024tenspiler}, the user provides a template to aid search.
While narrowly effective, such heuristics limit portability and are a limit to generalization.

\begin{figure}
  \includegraphics[width=\textwidth]{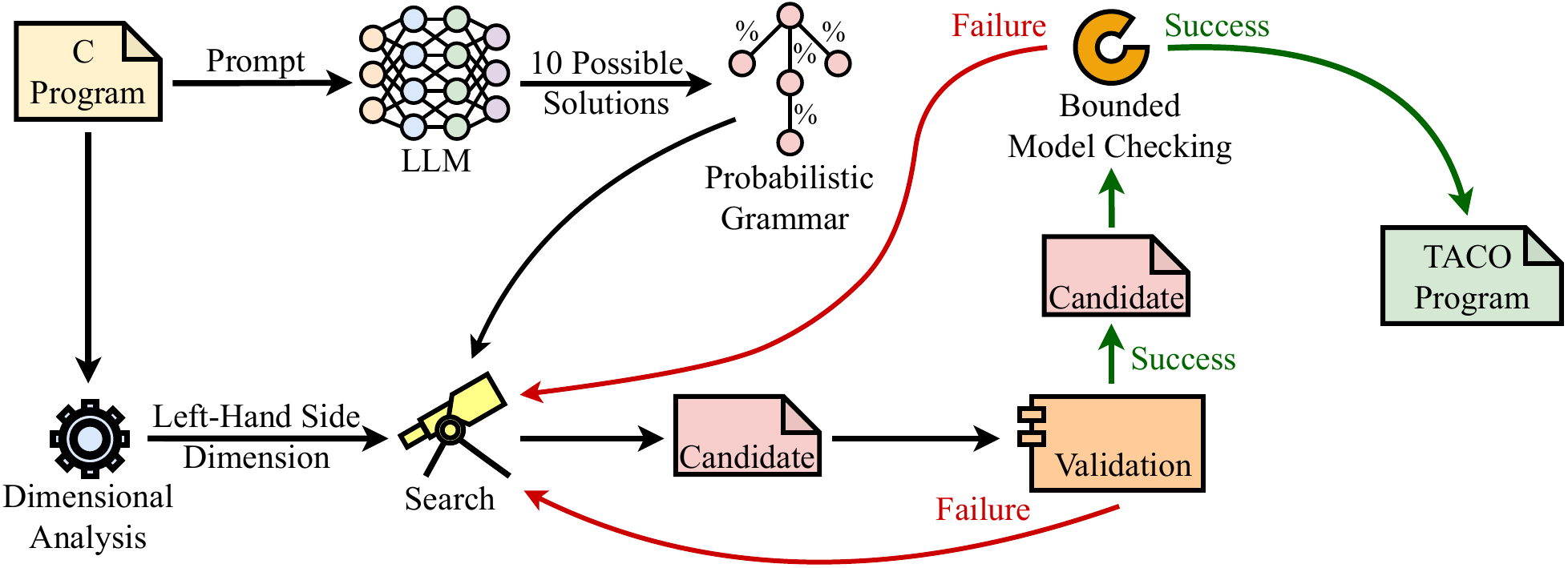}
  \caption{Overview of \sys. We query the LLM to provide 10 possible solutions in TACO that are equivalent to the input code C. Based on the LLM response, we build a probabilistic grammar and enumerate the space of template programs described by the said grammar. We validate a candidate using I/O examples, and if it passes all tests, we proceed to verification to prove equivalence with the original C implementation. The input code is also analyzed to predict the dimensionality of the left-hand side tensor in the solution.}  
  \Description{Overview of \sys.}
  \label{fig:overview}
\end{figure}

A completely different approach is to use neural machine translation based on 
large language models (LLMs). They have proved highly successful with a number of program generation tasks \cite{tree2tree,Drissi2018}.
They are fast and scale with program complexity, but unfortunately are inaccurate.
What we would like is to combine the power of LLMs with the accuracy of synthesis.

This paper explores a novel combination of LLMs and program synthesis.
It uses an LLM to suggest a number of possible 
solutions. It then builds a probabilistic grammar of templates, based on the proposed solutions, and then uses this grammar to drive an enumerative search of grammar templates.
Our approach, termed \sys (Synthesis of Tensor Algebra Guided by Grammars) is able to outperform all existing approaches. It achieves $99\%$ lifting accuracy on a pre-existing large-scale benchmark suite of dense tensor algebra and is able to do this without any pre-wired heuristics.

\subsection{Contributions}
This paper makes the following contributions:
\begin{itemize}
    \item Two novel synthesis algorithms that combine LLM guesses and program synthesis to scalably lift dense tensor code.     
    \item A large-scale evaluation of state-of-the-art tensor program lifting.
    \item Greater coverage than existing techniques.
\end{itemize}

\section{Motivation}
\label{sec:tensor-algebra}

The tensor DSL we target in our paper is TACO \cite{Kjolstad2017}.
Whilst TACO may be best known for sparse computation, it also gives superior performance over dense code on multicores and GPUs.
Although  recent work has tackled {\em matching} certain sparse
computation to specific high-performance APIs~\cite{SpEQ,ginsbach2020automatically}, this paper focuses solely on
{\em lifting} legacy dense tensor computation to a high-level programming language~\cite{kjolstad2017taco}. 

TACO syntax is based on Einstein summation (einsum) notation, a language for representing linear algebra operations using indexing expressions.
The TACO compiler takes a tensor expression as input and generates highly optimized kernels. 
TACO-generated code exploits the parallel nature of both dense tensor algebra operations and multi-core/GPU architectures. 
It uses domain-specific knowledge to optimize and auto-parallelize, which related work has reported results in average speedups of $1.8\times$ and $24.1\times$ over the original program on CPUs and GPUs, respectively~\cite{c2tatco}.

\textbf{\emph{Einsum notation:}}
TACO supports einsum notation, 
as do other
frameworks including PyTorch \cite{paszke2019pytorch}, Halide \cite{ragan2013halide}. While these alternatives may support a larger set of operations, targeting TACO
allows a direct apples-to-apples comparison against prior work \cite{c2tatco,Qiu2024}.

Einsum expressions consist of a sequence of indexing variables, each one representing an iterator over a different tensor dimension. The traditional einsum notation expresses tensor multiplication and implicit summation on the indices that are absent in the output tensor. TACO uses an extended version of the original notation that also supports subtraction and division. Unlike other einsum-based frameworks such as the NumPy~\cite{harris2020array} einsum API, the tensors in TACO programs must be explicitly declared. Figure~\ref{bnf:td} shows the TACO grammar addressed in this paper.

\textbf{\emph{Problem statement:}}
Formally, given a legacy program $\src$, written in a low-level language such as C, $\sys$ aims to find an equivalent program $\target$ written in TACO, that meets the specification 
$$\forall \inputs.  \src(\inputs)=\target(\inputs),$$
where $\inputs$ is a vector of input arguments. That is, $\target$ produces the same output as $\src$ on all possible inputs.

\subsection{Example}
\label{sec:example}

The synthesis task that \sys solves is to synthesize a TACO program $\target$ such that $\forall \inputs. \src(\inputs) = \target(\inputs)$, where $\inputs$ is a vector of inputs, in this case \texttt{Mat1, Mat2, Result}. We now illustrate this approach on the example input C program shown in Figure~\ref{query:benchmark}.

Given this program, \sys first queries a large language model to ask for a set of candidate solutions. The prompt template we use is shown below in Prompt~\ref{prompt}. This gives us the set of candidate solutions shown in Response~\ref{response:llm}.

\sys then learns a probabilistic context-free grammar that captures this set of solutions as templates.

We describe how we learn this grammar in Section~\ref{sec:pcfg}. The grammar in Figure~\ref{fig:example-template} shows probabilities for each production rule in parentheses. Each tensor and constant in the grammar is treated as a symbolic variable that can later be replaced when the template is instantiated.

We use a weighted A$^*$ search to explore the space of the grammar, inspired by work in the literature~\cite{LLM-SYGUS,euphony}, enhanced with penalty functions that penalize (partial or complete) templates that fail to adhere to syntactic constraints. When a complete template is found, this is passed to a template validator, which searches for all possible instantiations of the template and evaluates them against a set of input-output examples. A valid template, in this instance, would be the template \texttt{a(i) = b(i,j) * c(j)}. This is instantiated to the concrete program \texttt{Result(i) = Mat1(i,j) * Mat2(j)}. We compile this TACO program using the TACO compiler into C code, and check with bounded model checking that the two pieces of C code are equivalent.
\begin{figure}[H]
\begin{eg}[]
void function(int N, int* Mat1, int* Mat2, int* Result){
    int* p_m1;
    int* p_m2;
    int* p_t;
    int  i, f;
    p_m1 = Mat1;
    p_t  = Result;
    for (f = 0; f < N; f++) {
        *p_t = 0;
        p_m2 = &Mat2[0];
        for (i = 0; i < N; i++)
            *p_t += *p_m1++ * *p_m2++;
        p_t++;
    }
}
\end{eg}
\caption{A C implementation of $\sum_{i=0}^{N-1} \texttt{Mat1}(f \times N + i) \cdot \texttt{Mat2}(i)$. The result is a dot product between the $f$-th row of \texttt{Mat1} and vector \texttt{Mat2}. The equivalent synthesized TACO expression is \texttt{a(i) = b(i,j) $*$ c(j)}.\label{query:benchmark}}
\end{figure}

\begin{prompt}[numbers=none,caption={The prompt requesting $10$ TACO expressions for a given C program. The
temperature we use is $1.0$, and the role is “\texttt{You are a scientific assistant that knows a
lot about transpilation}”. },label={prompt}]
You are a scientific assistant that knows a lot about transpilation. Translate the following C code to an expression in the TACO tensor index notation. The expression must be valid as input to the taco compiler. Return a list with 10 possible expressions. Return the list and only the list, no explanations. 
{the input C program}
\end{prompt}

\begin{response}[caption={LLM-generated candidate solutions for matrix product computations based on the implementation in Figure~\ref{query:benchmark}. Displayed are a subset of the $10$ generated solutions, trimmed for brevity.},label={response:llm}]
r(f) = m1(i, f) * m2(f)
Result(i) = Mat1(i,f)*Mat2(f)
Result(i) := Mat1(f,i) * Mat2(i)
Result(f) = sum(f, mat1(f, i) * mat2(i))
\end{response}

\begin{figure}[H]
\begin{bnf}[]
<PROGRAM>      ::= <TENSOR1> EQ <EXPR> (1)
<TENSOR1>      ::= "a(i)" (1)
<EXPR>         ::= <TENSOR> (0) | <CONSTANT> (0) | <EXPR> <OP> <EXPR> (1)
<OP>           ::= ADD (0.2) | SUB (0) | MUL (0.8) | DIV (0)
<TENSOR>       ::= "b(i,j)" (0.2) | "b(j, k)" (0.1) |  $\ldots$
\end{bnf}
\begin{nobnf}[numbers=none]
             ::= "c(i)" (0.3) | "c(j)" (0.2) | "c(k)" 
\end{nobnf}
    \caption{A probabilistic context-free grammar template.}
    \label{fig:example-template}
\end{figure}

\section{Overview of \sys}
Lifting to tensor DSLs is a challenging problem for program synthesis, and existing enumerative techniques are capable of accurate translation but rely on hand-written heuristics in order to scale. In contrast, highly scalable machine-learning-based approaches like language models fail to give accurate translations due to the complexity of the benchmarks. 
The key insight behind \sys is that we can achieve the best of both worlds by using an LLM to \emph{learn} the heuristics for an enumerative solver.


To that end, \sys, as shown in Figure ~\ref{fig:overview}, implements a multi-staged hybrid synthesis approach. \begin{description}
    \item[\ding{172}] First, we construct a prompt based on the input C code and ask the LLM to propose $10$ translated solutions. 
    \item[\ding{173}] We proceed to construct a probabilistic grammar, which represents the space of solutions in the form of templates.
    \item[\ding{174}] We then search this space of templates with a two-stage enumerative search: first, we search the space of templates with a search inspired by A$^*$, then, given a template, we search for a valid completion of the template against a set of input-output examples. 
    \item[\ding{175}] If a completed template is found that satisfies all input-output examples, we perform bounded verification with a bounded model checker to validate that the completed template is equivalent to the original C code. If it fails verification, we return to the template enumeration stage. 
\end{description}

\section{Learning a Grammar of Templates}
\label{sec:pfcg}
The first step of \sys uses a large language model to generate a set of candidate solutions for the synthesis problem in hand, using the prompt shown in Section~\ref{sec:example}. This gives us a set of candidate solutions, $P$. We ask for $10$ solutions, but we parse in as many solutions as the LLM gives us (which is sometimes more than $10$) and discard any syntactically incorrect solutions. 

Given a set of incorrect candidate solutions from the LLM, we hypothesize that, even though none of the candidate solutions were precisely correct, the correct solution is likely to lie in the neighborhood of the LLM's guesses. To that end, we characterize this neighborhood using a probabilistic grammar of templates. We use a context-free grammar but note that, in principle, any probabilistic model that characterizes the neighborhood of guesses could be used. First, let us define some of the preliminaries we will need for this section. 

\subsection{Preliminaries}
\begin{definition}[Context-Free Grammar, CFG]\label{def:cfg} A context-free grammar (CFG) is a 4-tuple, $$\grammar = \langle \nt, \T, \rules, \start \rangle. $$ $\nt$ is a finite set of non-terminal symbols. $\T$ with $\T \cap \nt = \emptyset$ is a finite set of terminal symbols. $\rules \subseteq \nt \times (\nt \cup \T)\kleeneStar$ is a finite set of production rules, where $\kleeneStar$ denotes the Kleene star. Each production rule in a context-free grammar is of the form $\alpha \rightarrow \beta$, where $\alpha$ is a symbol in the set of non-terminals $\nt$, and $\beta$ is a string composed of symbols from $(\nt \cup \T)\kleeneStar$.  $\start \in \nt$ is the start symbol of the grammar $\grammar$. 

Given a context-free grammar $\grammar = \langle \nt, \T, \rules, \start \rangle$, with $x, y \in (\nt \cup \T)\kleeneStar$ and a rule $(\alpha \rightarrow \beta) \in \rules $, we write $x \Rightarrow_G\kleeneStar y$ if there exist strings $u, v \in (\nt \cup \T)\kleeneStar$ such that $x = u\alpha v$ and $y = u\beta v$, and we write $x \Rightarrow\kleeneStar y$ if either $x = y$ or $x \Rightarrow x_1 \Rightarrow \dots \Rightarrow x_n \Rightarrow y$ for $n \geq 0$. The language generated by $\grammar$, denoted $\languageOfGrammar$, is the set of all strings over $\T$ that can be derived from the start symbol $\start$ by applying a sequence of production rules from $\rules$. Formally, we define $$\languageOfGrammar = \{ \sigma \in \Sigma\kleeneStar \mid S \Rightarrow_G\kleeneStar \sigma \}.$$

\end{definition}

\begin{definition}[Weighted Context-Free Grammar, CFG]\label{def:wcfg}
A weighted context-free grammar ($wCFG$) is a 5-tuple, $$ wCFG= \langle \nt, \T, \rules, \start, \weight \rangle,$$ contains a $G = \langle \nt, \T, \rules, \start \rangle$, and a weighted function $\weight:R \rightarrow \mathbb{R}^+$ maps each production rule to a positive real number. 
\end{definition}

\begin{definition}[Probabilistic Context-Free Grammar, pCFG]\label{def:pcfg}
A probabilistic context-free grammar, $$pCFG = \langle \nt, \T, \rules, \start, \prob \rangle,$$ is a specific instance of $wCFG$, with the additional constraint that the weights of the production rules expanding each non-terminal must sum to one. The probability function $\prob : \rules \rightarrow [0, 1]$ in $pCFG$ assigning a probability $\prob(\alpha \rightarrow \beta)$ to each rule. The probabilities must satisfy 
$$
\forall \alpha \in \nt. \sum_{\beta : (\alpha \rightarrow \beta) \in \rules} \prob(\alpha \rightarrow \beta) = 1.$$
\end{definition}

\subsection{Constructing a Grammar of Templates}

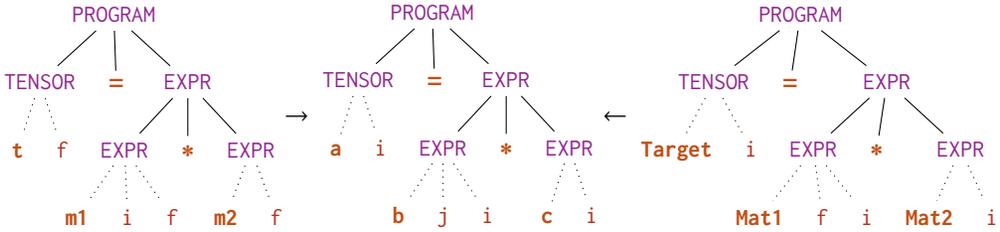
\begin{figure}[H]
    \centering
    \small
    \begin{minipage}{0.27\textwidth}
        \begin{forest}
[\texttt{\color{Plum}PROGRAM}
    [\texttt{\color{Plum}TENSOR}
            [\fontseries{b}\selectfont{\texttt{\textcolor{Bittersweet}t}}, edge=dotted]
            [\texttt{\textcolor{BrickRed}f}, edge=dotted]
    ]
    [{\color{Bittersweet}$\boldsymbol=$}]
    [\texttt{\color{Plum}EXPR}
        [\texttt{\color{Plum}EXPR}
                [\fontseries{b}\selectfont{\texttt{\textcolor{Bittersweet}{m1}}}, edge=dotted]
                [\texttt{\textcolor{BrickRed}i}, edge=dotted]
                [\texttt{\textcolor{BrickRed}f}, edge=dotted]
        ]
        [\normalsize{$\color{Bittersweet}\boldsymbol*$}]
        [\texttt{\color{Plum}EXPR}
                [\fontseries{b}\selectfont{\texttt{\textcolor{Bittersweet}{m2}}}, edge=dotted]
                [\texttt{\textcolor{BrickRed}f}, edge=dotted]
        ]
    ]
]
        \end{forest}
    \end{minipage}
    $\rightarrow$
    \begin{minipage}{0.27\textwidth}
        \begin{forest}
[\texttt{\color{Plum}PROGRAM}
    [\texttt{\color{Plum}TENSOR}
            [\fontseries{b}\selectfont{\texttt{\textcolor{Bittersweet}a}}, edge=dotted]
            [\texttt{\textcolor{BrickRed}i}, edge=dotted]
    ]  
    [{\color{Bittersweet}$\boldsymbol=$}]
    [\texttt{\color{Plum}EXPR}
        [\texttt{\color{Plum}EXPR}
                [\fontseries{b}\selectfont{\texttt{\textcolor{Bittersweet}b}}, edge=dotted]
                [\texttt{\textcolor{BrickRed}j}, edge=dotted]
                [\texttt{\textcolor{BrickRed}i}, edge=dotted]
        ]
        [\normalsize{$\color{Bittersweet}\boldsymbol*$}]
        [\texttt{\color{Plum}EXPR}
                [\fontseries{b}\selectfont{\texttt{\textcolor{Bittersweet}c}}, edge=dotted]
                [\texttt{\textcolor{BrickRed}i}, edge=dotted]
        ]
    ]
]
        \end{forest}
    \end{minipage}
    $\leftarrow$
    \begin{minipage}{0.27\textwidth}
        \begin{forest}
[\texttt{\color{Plum}PROGRAM}
    [\texttt{\color{Plum}TENSOR}
            [\fontseries{b}\selectfont{\texttt{\textcolor{Bittersweet}{Target}}}, edge=dotted]
            [\texttt{\textcolor{BrickRed}i}, edge=dotted]
    ]  
    [{\color{Bittersweet}$\boldsymbol=$}]
    [\texttt{\color{Plum}EXPR}
        [\texttt{\color{Plum}EXPR}
                [\fontseries{b}\selectfont{\texttt{\textcolor{Bittersweet}{Mat1}}}, edge=dotted]
                [\texttt{\textcolor{BrickRed}f}, edge=dotted]
                [\texttt{\textcolor{BrickRed}i}, edge=dotted]
        ]
        [\normalsize{$\color{Bittersweet}\boldsymbol*$}]
        [\texttt{\color{Plum}EXPR}
                [\fontseries{b}\selectfont{\texttt{\textcolor{Bittersweet}{Mat2}}}, edge=dotted]
                [\texttt{\textcolor{BrickRed}i}, edge=dotted]
        ]
    ]
]
        \end{forest}
    \end{minipage}
    \hspace{0.1\textwidth} 

    \caption{Expression standardization. We omit part of the derivation for brevity.}
    \label{fig:standardization}
\end{figure}

Capturing the search space as a grammar of templates rather than complete TACO programs has two advantages: first, in comparison to using the full TACO grammar, it reduces the search space that the enumerative synthesis process has to search; and second, it allows us to group semantically equivalent but syntactically different candidate expressions together as one when constructing the weighted context-free grammar. 
For example, expressions like \texttt{\color{Sepia}{\small 1} \color{black}t(f) = m1(i, f) $*$ m2(f)} and \texttt{\color{Sepia}{\small 3} \color{black}Target(i) $\coloneqq$ Mat1(f,i) $*$ Mat2(i)} in LLM Response~\ref{response:llm}, are equivalent in structure (we use preprocessing to swap $\coloneqq$ to $=$ before parsing) shown in Figure~\ref{fig:standardization}, yet they would yield different terminal rules in the full grammar due to variations in notation.

Given a set of candidate solutions, the first step is to construct a grammar of \emph{templates} that captures the full set of solutions. The full grammar for TACO programs, $G_{TACO}$ is shown in Figure~\ref{bnf:td}. A TACO program is any program in $\languageOfGrammar(\grammar_{TACO})$.

\begin{definition}[Templates]
We define a \emph{TACO template} $\template$ to be any string obtained by taking a program $p \in \languageOfGrammar(\grammar_{TACO})$ and replacing all tensors with symbolic tensor variables, denoted $\tensor_1, \tensor_2, \ldots$ and all constants with constant symbols $\const_1, \const_2, \ldots$ 
\end{definition}

Given a set of candidate programs from the LLM, $P$, we aim to find a grammar  $G_{\template}$ that contains a set of templates $\templateset$ that allow us to generate all programs $p \in P$. 
A substitution $S = (\tensor_1 \mapsto s^1_1, \tensor_2 \mapsto s^1_2, \ldots, \const_1 \mapsto s^c_1 )$ is a mapping from symbolic tensor variables and constant symbols to terminal symbols in the grammar $\grammar_{TACO}$.
A template \emph{generates} a program if $\exists S. \template.\{S\} = p$, where $\template.{S}$ indicates the result of replacing all occurrences of $\tensor_1$ with $s^t_1$, and $\tensor_2$ with $s^t_2$ etc in $t$. 
Thus, our requirement on our grammar $G$ is the following constraints: first that 
$$\forall \template. (\template \in \templateset) \implies \template \in \languageOfGrammar(G),$$ and second that 
     $$\forall p \in P. \exists \template \in \templateset \wedge \exists S.\, \template.\{S\} = p.$$
     
This is obviously trivially satisfied by the complete grammar of TACO programs, yet it is undesirable as it has not reduced our search space. We also aim to create a grammar that is as small as possible, whilst avoiding over-fitting, and we attempt to optimize this trade-off by the method of construction described in the following section.

\begin{figure}[H]
\begin{nobnf}
<PROGRAM>      ::= <TENSOR> EQ <EXPR>
<TENSOR>       ::= <IDENTIFIER> | <IDENTIFIER> "(" <INDEX-EXPR> ")"
<EXPR>         ::= <TENSOR> | <CONSTANT> | "(" <EXPR> ")" | SUB <EXPR> |
\end{nobnf}
\begin{nobnf}[numbers=none]
                 <EXPR> ADD <EXPR> | <EXPR> SUB <EXPR> |
                 <EXPR> MUL <EXPR> | <EXPR> DIV <EXPR>
\end{nobnf}
\begin{bnf}[firstnumber=4]
<INDEX-EXPR>   ::= <INDEX-VAR> | <INDEX-VAR> "," <INDEX-EXPR>
<INDEX-VAR>    ::= "i" | "j" | "k" | "l"
<IDENTIFIER>   ::= <LETTER> ( <LETTER> | <INTEGER> )*
<CONSTANT>     ::= <INTEGER>
<INTEGER>      ::= <DIGIT>+
<LETTER>       ::= "a" | "b" | ... | "z" | "A" | "B" | ... | "Z"
<DIGIT>        ::= "0" | "1" | "2" | ... | "9"
\end{bnf}
\caption{The grammar for TACO expression in Extended Backus–Naur form, defining the syntax for tensor expressions, identifiers, constants, and basic arithmetic expressions. The $\kleeneStar$ symbol denotes Kleene star, indicates zero or more repetitions of the preceding element, while $\kleenePlus$ denotes Kleene plus, requires one or more occurrences. \label{bnf:td}}
\end{figure}

\subsubsection{Templatized Candidate Solution}

We obtain the grammar $G_{\template}$ by first inferring a template for each solution in $P$.
The first step involves parsing each $p\in P$ into an Abstract Syntax Tree (AST), a structured representation that captures the hierarchical organization of the expression. The AST organizes operations, tensor identifiers, and indices as distinct nodes, allowing systematic traversal and manipulation. For example, consider the expression \texttt{\color{Sepia}{\small 1} \color{black}t(f) = m1(i, f) $*$ m2(f)} in Response~\ref{response:llm} can be parsed as the left tree in Figure~\ref{fig:standardization}.
We then transform the AST in three stages: \textit{Tensor Templatization}, \textit{Index Standardization}, and  \textit{Constant Templatization}.

\paragraph{Tensor Templatization}
We replace each tensor name in the expression with a symbolic tensor variable. From hereon, we use \texttt{a}, \texttt{b}, \texttt{c}, as the symbolic tensor variables $\tensor_1, \tensor_2, \tensor_3, \ldots$, to align with the variable names in the code examples. 
The identifiers are assigned in alphabetical order—starting with \texttt{a} for the left-hand side tensor and using \texttt{b}, \texttt{c}, \texttt{d}, ... sequentially for tensors on the right-hand side, based on their order of first appearance. 
The expression \texttt{t(f) = m1(i, f) $*$ m2(f)} will be transformed into \texttt{a(f) = b(i,f) $*$ c(f)} by this step. 

\paragraph{Index Standardization}
The index standardization step ensures that each tensor expression in the grammar uses a consistent set of index variables, irrespective of the original indices in the input expression. 
Each unique index variable encountered in an expression is mapped to the next available symbol from the canonical set $\{ \texttt{i}, \texttt{j}, \texttt{k}, \texttt{l}\}$ in alphabetical order. The expression \texttt{a(f) = b(i,f) $*$ c(f)} will be transformed to \texttt{a(i) = b(j,i) $*$ c(i)} by this step, as shown in the middle in Figure~\ref{fig:standardization}. The index variables do not need to be replaced by template instantiation as they are local variables to the program. 

\paragraph{Constant Templatization}
Any constants in the candidate solutions are replaced with a symbolic constant \texttt{Const}. The template instantiation step will instantiate from a list of constants found in the input source code. 

\subsubsection{Refining the Grammar}

Given the templatized candidate solutions $\templateset$, we wish to construct a probabilistic grammar that represents the space of these solutions, without substantially over-fitting. The first step is to construct a context-free grammar that defines this set of templates. 

We start with the base grammar of TACO programs, shown in Figure~\ref{bnf:td}. We then restrict the set of tensor names to be the names we have chosen as symbolic tensor variables and constants, namely $a$, $b$, $c$, $\dots$ and $const$, and also limit the set of index variables to be $i,j,k,l,\ldots$. In theory, this permits $26$ tensor IDs and $4$ index variables, because one can always infer whether a variable is an index or a tensor identifier by context. 
In practice, we never need this many, and searching a space that includes up to $26$ $4$-dimensional tensors is obviously impractical. This section addresses how we initially reduced this search space. Namely, by predicting the dimensions of the tensors in order to refine the grammar. 

\subsubsection{Predicting Tensor Dimensions}
To accurately predict tensor dimensions for a given program, we combine insights from a language model (LLM) with static code analysis. 
Static analysis is used to predict the left-hand side (LHS) tensor dimensions of an expression, while the LLM is used to predict dimensions for the right-hand side (RHS). Static analysis, by analyzing the source code, can determine precisely the dimensions for the LHS tensor, but cannot do the same for the RHS, so we fall back on heuristics learned by the LLM for the RHS.

\begin{definition}[Dimension list] 
We define a dimension list $L_\template$ to be a list of integers $(d_1, d_2, d_3,\ldots)$ where $d_i$ is the dimension for the $i^{th}$ unique tensor in the template $\template$. We use $L[i]$ to indicate accessing the $i^{th}$ element of the dimension list $L$, and $|L|$ to indicate the length of the list $L$. We list the dimensions of constants and variables as $0$. 
\end{definition}
 For example, the list $[1,2,4]$ indicates that the first tensor in the expression has $1$ dimension, the second tensor has $2$ dimensions, and the third tensor has $4$ dimensions. 

\paragraph{Dimension Prediction for RHS Tensors Using LLM}
Given a set of templatized solutions $\{\template_1, \ldots \template_{10}\} \in \templateset$, generated by the LLM, we compute the dimension list for each candidate solution, giving a set of lists $L_{\templateset} = \{L_{\template_1}, \ldots, L_{\template_{10}}\}$. We then filter this list by length, and remove any list that has a length less than the maximum length, giving the filtered set 
$$L^*_\templateset = \{l \in L_{\templateset} \,\mid \,|l| \geq |L_{\template_i}|\; \forall L_{\template_i}\in L_P\templateset)\}.$$

Finally, we return the list that appears most frequently in the filtered set, i.e., 
$$L_\template = \text{arg}\,\max\limits_{l}\,  | l \in L^*_{\templateset} |. $$ 
From here on, we denote this predicted list $L$, and refer to this as the predicted dimension list.

\paragraph{Integrating Static Analysis for LHS Tensors: }
\label{sec:lhs-tensor-dimension}
We use static program analysis to examine the original program AST and predict the LHS dimension. 
We apply a dataflow analysis to recover the dimensions in the array accesses to recover the original dimensionality. For standard array accesses, e.g., $a(i,j)$, we simply count the number of variables used to index the base pointer. However, it is common that C programs access multi-dimensional elements using affine linear expressions on index variables. In such cases, we use array delinearization \cite{OBoyle2002} to recover the standard array access form and predict the dimensionality by counting the number of indexing variables. 
Additionally, some applications use explicit pointer arithmetic to iterate over arrays. We implement array recovery \cite{Franke2003} to retrieve array access expressions from pointers and then apply delinearization and analyze the indexing expression. In case the output variable is not accessed through any memory indexing operation, we assume it is a scalar and predict zero-dimensionality. 

As the left-hand side tensor necessarily appears first in the expression, we replace $L[1]$ with the predicted dimension for the first tensor from the static analysis.

\subsubsection{Generating the Context-Free Grammar}
Given a dimension list, we wish to generate a grammar that ensures we only enumerate combinations of indices required to make all possible tensor expressions that match the predicted dimensions (or, at least reduce this space as far as possible without increasing the complexity of the grammar significantly). 
For instance, if the dimension list is $[0, 1, 3]$, and at least one of the predicted solutions is \texttt{a = b(i) + c(i,j,k)}, we will modify the grammar to fix the production rule for the first tensor to restricted to \texttt{a}, and ensure the grammar can enumerate \texttt{b(i)} and \texttt{c(i,j,k), c(i,k,j), c(j,i,k), c(k,i,j), c(j,k,i), c(k,j,i)} for any remaining tensor that appear in the expression. 
This permits all possible combinations of indexing of the tensors of the predicted dimensions, allowing the index used for \texttt{b(i)} to be repeated in any position when indexing \texttt{c}.

Formally, we define this grammar generator as a set of constraints reasoning over the predicted dimension list $L$ and the set of templates $\templateset$, which, if true, indicate that the production rule should be included in the grammar. We use \texttt{[c]$r_i$} to denote that a production $r_i$ is included within the grammar if \texttt{c} is true.
$i(P)$ denotes the number of unique index variables in the set of programs $\templateset$. Rules without a constraint automatically appear in the grammar.
\begin{bnf}[]
         <PROGRAM> ::= <TENSOR1> EQ <EXPR>
         <EXPR>    ::= <TENSOR> | <EXPR> <OP> <EXPR> 
         <OP>      ::= ADD | SUB | MUL | DIV
[$L[1]=0$]<TENSOR1>        ::= "a"
[$L[1]=1$]<TENSOR1>        ::= "a(i)" 
[$L[1]=2$]<TENSOR1>        ::= "a(i,j)"
            $\ldots$
[$L[2]=0$]<TENSOR>         ::= "b" | "Const" 
[$L[2]=1$]<TENSOR>         ::= "b(i)" | [$i(P)>1$]"b(j)" | [$i(P)>2$]"b(k)" | 
            $\ldots$ 
[$L[2]=2$]<TENSOR>         ::= "b(i,j)" | "b(j,i)" | [$i(P)>2$]"b(i,k)" | $\ldots$ 
            $\ldots$
[$L[3]=0$]<TENSOR>         ::= "c" | "Const" 
[$L[3]=1$]<TENSOR>         ::= "c(i)" | [$i(P)>1$]"c(j)" | [$i(P)>2$]"c(k)" | $\ldots$  
            $\ldots$
\end{bnf}

For every element in the dimension list, we add a new tensor id, \texttt{a, b, c \ldots}, and index it with the number of index variables that correspond to the element in the dimension list. For an element $i$, where $L[i] = n$, and for a set of candidate programs where $i(\templateset) = m$, we add a production rule for every possible way of choosing $n$ indices from $m$ index variables. We then remove any production rules that could not be used parsing any candidate $\template \in \templateset$, e.g., if no template in $\templateset$ contains a 2-dimension tensor indexed with the same index variable twice, we will remove \texttt{b(i,i)}.
An example generated template grammar is shown in Figure~\ref{bnf:template}.

\begin{figure}[H]
\begin{bnf}[]
<PROGRAM>   ::= <TENSOR1> EQ <EXPR>
<TENSOR1>   ::= "a(i)" 
<EXPR>      ::= <TENSOR> | <CONSTANT> | <EXPR> <OP> <EXPR> 
<TENSOR>    ::= "b(i,j)" | "b(j, i)" | "b(i, k)" | $\ldots$
\end{bnf}
\begin{nobnf}[numbers=none]
              "c(i)" | "c(j)" | "c(k)" 
\end{nobnf}
\begin{bnf}[firstnumber=5]
<CONSTANT>  ::= "Const"
<OP>        ::= ADD | SUB | MUL | DIV 
\end{bnf}
\caption{An example generated template grammar, for the dimension list $[1,2,1,0]$, with a maximum of $3$ unique indices appearing in the candidate LLM solutions. \label{bnf:template}}
\end{figure}

\subsection{Assigning Probabilities to the Grammar}
\label{sec:pcfg}
We now wish to assign a probability to each production rule in the grammar of TACO templates according to their frequency in the left-most derivations of the candidate solutions. 
Let us define a derivation as follows:
\begin{definition}[Derivations] Given a context-free grammar $\grammar$ as previously defined, and a sentence $s \in \languageOfGrammar(\grammar)$, the derivation of $s$ from $S$ is a sequence of rules such that $S \xrightarrow{r_0} s_1 \xrightarrow{r_1}\ldots s_n \xrightarrow{r_n} s$ and $r_0 \ldots r_n \in \rules$. We denote the derivation of $s$ by the sequence of rules $r_0, \ldots r_n$ as $D_s = \{r_0, \ldots r_n\}$. The left-most derivation is a derivation such that all rules expand the left-most non-terminal symbol in the sentential form.
\end{definition}

Given a set of templatized solutions $\templateset \in \languageOfGrammar(G_{template})$, we calculate a weight for each rule $r_i \in R$ as the number of times that rule appears in the left-most derivations of the programs. That is, 
\begin{equation*}
\weight[r_i]  = \sum_{\template_i \in \templateset} |r_i| \in D_{\template_i},
\label{eq:weightupdate}
\end{equation*}
where $|r_i|$ is the number of times $r_i$ appears in the derivation.
These weights reflect the usage frequency of each tensor with specific indices in the expressions provided by the LLM.
Note that, for any production rules used to replace the tensor nonterminal symbols, e.g., \texttt{\color{Plum}{1DTENSOR}}, which do not appear in any of the candidate solutions, we assign a default weight of $1$.
This assignment ensures that these combinations are considered during the synthesis process with a lower priority.

Using the weights calculated for both operators and tensors, we construct the corresponding probabilistic Context-Free Grammar ($pCFG_{\template}$) by normalizing the weights into probabilities.
For each non-terminal symbol $\alpha$, the probability of applying the production rule $\alpha \rightarrow \beta$ is given by:

\begin{equation*}
    \prob[\alpha \rightarrow \beta] = \frac{\weight[\alpha \rightarrow \beta]}{\sum_{\gamma} \weight[\alpha \rightarrow \gamma]},
\end{equation*}
where the sum in the denominator is over all production rules where $\alpha$ appears on the left-hand side, and $\gamma$ may be any string of nonterminal and terminal symbols.

\section{Searching the Template Space}
We present two algorithms for searching the space of TACO templates. The first is based on a weighted \astar search in the literature~\cite{euphony,LLM-SYGUS}, which searches the grammar of TACO templates in a top-down manner. We extend this algorithm to incorporate a penalty score that accounts for known syntactic constraints on the target solution. The second is an adapted version of \astar, which combines bottom-up search with \astar heuristics.  

\subsection{Top-Down Weighted \astar}
\label{sec:top-down-weighted-astar}
Algorithm~\ref{algo:td} outlines the top-down weighted \astar search with penalties. The search operates over a probabilistic Context-Free Grammar derived from large language model (LLM) outputs.
It maintains a queue of partial templates, represented as abstract syntax trees, which we can think of as the frontier of its search. This initially contains just the start symbol of the grammar.  At each iteration, it must determine which of the partial templates should be further explored, which it does based on the cost of the path taken to reach those partial templates and an estimate of the cost required to extend the path all the way to the goal. The goal is, ultimately to find a complete template that we believe is likely to satisfy the specification, that is a template can generate a program $\target$ such that $\forall \inputs. \target(\inputs) = \src(\inputs)$.

Thus, when choosing which partial template to further explore, the algorithm chooses the template with the minimum $f(x)$, where $x$ is the partial template, defined as:
\begin{equation*}
f(x) = c(x) + g(x) + \Penalty(x),
\label{eq:total_cost}
\end{equation*}
where $c(x)$ calculates the accumulated cost from the start node $S$ to current node $x$. 
$g(x)$ is the heuristic estimate of the minimal cost to complete the expression from $x$ to a goal node (a template we believe is likely to satisfy the specification) and $\Penalty(x)$ is a penalty term for expressions that violate domain-specific syntactic constraints. These are calculated as follows:

\noindent\textbf{The accumulated cost $c(x)$} is calculated as the sum of the costs of the production rules applied along the path to $x$:
\begin{equation*}
c(x) = \sum_{r_i \in D_x} -\log_2 \prob[r_i],
\end{equation*}
where $D_x$ is the sequence of production rules used to reach the node $x$ and $\prob[r_i]$ is the probability of production rule $r_i$.
This cost function transforms probabilities into additive costs, suitable for the \astar search.

\noindent\textbf{The heuristic function $g(x)$} estimates the minimal additional cost required to complete the partial expression at node $x$ to a full expression. It is defined as:

\begin{equation*}
g(x) = \begin{cases}
    0 & \text{if } x \in \T^*, \\
    -\sum_{x_i \in \nt} \log_2 h(x_i) & \text{otherwise},
\end{cases}
\label{eq:heuristic_g}
\end{equation*}

where $x_i$ are the non-terminal symbols in the partial expression $x$, and $h(\alpha)$ is the maximal probability of deriving any terminal string from non-terminal $\alpha$.
The value $h(\alpha)$ is defined recursively for each non-terminal $\alpha$:

\begin{equation*}
\forall \alpha \in \nt, \quad h(\alpha) = \max_{\alpha \rightarrow \beta \in \rules} \left( \prob[\alpha \rightarrow \beta] \times \prod_{\beta_i \in \beta} h(\beta_i) \right),
\label{eq:h_alpha}
\end{equation*}

with the base case $h(\alpha) = 1, \;\text{if } \alpha \in \T.$
This equation represents the maximal probability of deriving a terminal string from $\alpha$, accounting for the probabilities of production rules and the maximal probabilities of its components.

\noindent\textbf{The penalty function $\Penalty(x)$} assigns additional costs to expressions that do not meet specific domain criteria. This function can be formalized as follows:
$$
\Penalty(x) = 
\begin{cases}
    \sum_{a \in A} \Penalty_{a}(x) & \text{if } x \text{ violates criterion } a, \\
    0 & \text{otherwise},
\end{cases}
$$
There are $5$ criteria $\{a_1, \ldots a_5\} \in A$, and their penalty scores are defined as follows (note that an infinite penalty score effectively means these expressions will never be considered):
\begin{itemize}
\item $\Penalty_{a_1}(x) = 10$, where $a_1$ is violated if the grammar includes a constant expression, the length of $x$ exceeds $3$, and $x$ either 1) contains fewer than $2$ tensors with index \texttt{i} or 2) lacks a constant expression. This penalty biases the search against expressions with multiple tensors but inadequate index variety or missing constants.
\item $\Penalty_{a_2}(x) = 100$, where $a_2$ is violated iff $x$ is not the same length as the length of the dimension list.
\item $\Penalty_{a_3}(x) = \infty$, where $a_3$ is violated if the tensor symbols in $x$ are not in alphabetical order by order of first appearance. This penalty rule avoids enumerating templates that are structurally identical, and therefore can be instantiated into identical sets of programs.
\item $\Penalty_{a_4}(x) = \infty$, where $a_4$ is violated if $x$ is a complete template (no non-terminal symbols), and repeatedly applies addition, subtraction, or division operations on the same tensor.
\item $\Penalty_{a_5}(x) = \infty$, where $a_5$ is violated if $x$ is a complete template (no non-terminal symbols), and employs fewer than half of the operations defined in the grammar.

\end{itemize}

\begin{algorithm}[h]
\caption{Top-Down Enumerator}
\label{algo:td}
\begin{algorithmic}[1]
\Procedure{Enumerate}{$pCFG_{\template}$}
    \State $Q \gets \{(0, pCFG_{\template}.S)\}$ \Comment{Initialize queue with start symbol of grammar}
    \While{$Q \neq \emptyset$}
        \State $(f, x) \gets Q.pop()$ \Comment{Remove template with minimal $f$}
        \If{$\text{depth}(x) > \text{maxDepth}$}
            \State \textbf{continue} \Comment{Skip if maximum depth exceeded}
        \EndIf
        \If{$x \in \Sigma^*$} \Comment{If no non-terminals remain in $x$}
            \State $S \gets \textsc{Validate}(x)$ \Comment{Try to instantiate $x$}
            \If{$S \neq \bot$}
                \If{\textsc{Verify}$(x.\{S\})$}
                 \State \Return $x.\{S\}$
            \EndIf
            \EndIf
        \EndIf
        \For {$x'$. s.t. $(x \xrightarrow{r} x' \wedge \,\, r \in pCFG_\template.R)$ } \Comment{Iterate over all possible expansions of $x$}
                \State $Q \gets Q \cup \{c(x') + g(x') + \Penalty(x'),  x'\}$
        \EndFor
    \EndWhile

    \State \Return Failure \Comment{Return Failure if no valid expression is found}
\EndProcedure
\end{algorithmic}
\end{algorithm}

\noindent \textbf{Search:} the search is shown in Algorithm~\ref{algo:td}. The algorithm keeps a queue of expressions in a queue, which you can consider to be the frontier of the search. At each iteration, it selects the expressions with the lowest total score $f$ from the queue.
If the expression is a complete expression, i.e., it contains only terminal symbols from the grammar, we then send this to the validation procedure described in Section~\ref{sec:validation}. 
If the expression is a partial expression, the leftmost non-terminal of the expression is expanded according to all applicable production rules in the grammar, creating a new template for each production rule. These new expressions are all added to the queue and the process is repeated. 

We set a depth limit of $6$, and if any expression exceeds this depth, it is discarded. We calculate depth as the depth of the maximum child in the abstract syntax tree, excluding index expressions, e.g., \texttt{b(i)} and \texttt{c(i,j)} are both expressions of depth 1, and \texttt{b(i) + c(i,j)} is an expression of depth 2.

\subsection{Bottom-Up Weighted \astar }
\label{sec:bottom-up-weighted-astar}
The algorithm presented in the previous section takes a top-down approach to enumerating through the search space. This has advantages over bottom-up search, namely that it is known to find longer programs faster than bottom-up search, which is biased towards shorter programs. Nevertheless, recent work has shown that guided bottom-up search can produce promising results~\cite{nadia}. To that end, we develop a new \astar inspired bottom-up search algorithm, which we term bottom-up \astar. 
The bottom-up search, shown in Algorithm~\ref{algo:bu} constructs expressions incrementally by starting with basic tensors and systematically combining them using operators, following a probabilistic context-free grammar. 
Again, the algorithm maintains a queue of expressions, and uses the same combination of cost, estimated cost to reach a goal state and the penalty function to determine which expression to expand first.

One key difference in the bottom-up search is the way we generate the template grammar. For the bottom-up search where the production rules only permit extending an expression by adding an operator and a new tensor to the end, effectively forcing the algorithm to enumerate programs shortest first. The grammar generator, given a predicted dimension list $L$, and a function $i(\templateset)$ which calculates the number of unique indices in $\templateset$, is shown below:
\begin{bnf}[]
       <PROGRAM>  ::= <TENSOR1> EQ <EXPR>
       <EXPR>     ::= <TENSOR2> <TAIL1>
       <TAIL1>    ::= EOL | [$|L|>2$] <OP> <TENSOR3> <TAIL2> 
       <TAIL2>    ::= EOL | [$|L|>3$] <OP> <TENSOR4> <TAIL3> 
\end{bnf}
\begin{nobnf}[numbers=none]
        $\ldots$
\end{nobnf}
\begin{bnf}[firstnumber=5]
       <OP>       ::= ADD | SUB | MUL | DIV  
[$L[1]=0$]<TENSOR1>      ::= "a"
[$L[1]=1$]<TENSOR1>      ::= "a(i)" 
[$L[1]=2$]<TENSOR1>      ::= "a(i,j)"
[$L[2]=0$]<TENSOR2>      ::= "b" | "Const"
[$L[2]=1$]<TENSOR2>      ::= "b(i)" | [$i(P)>1$]"b(j)" | [$i(P)>2$]"b(k)" 
[$L[2]=2$]<TENSOR2>      ::= "b(i,j)" | "b(i,j)" | "b(j,i)" | 
\end{bnf}
\begin{nobnf}[numbers=none]
                    [$i(P)>2$]"b(i,k)" | $\ldots$
\end{nobnf}
\begin{bnf}[firstnumber=12]
[$L[3]=0$]<TENSOR3>       ::= "c" | "Const" 
[$L[3]=1$]<TENSOR3>       ::= "c(i)" |[$i(P)>1$]"c(j)" |[$i(P)>2$]"c(k)" | $\ldots$ 
$\ldots$
\end{bnf}

An example of generated grammar is shown in Figure~\ref{bnf:bu}.
Weights and probabilities over the grammar are then calculated as described in Section~\ref{sec:pcfg}. 

\begin{figure}[H]
\begin{bnf}[]
<PROGRAM>    ::= <TENSOR1> EQ <EXPR> 
<TENSOR1>    ::= "a" 
<EXPR>       ::= <1DTENSOR> <TAIL1> 
<TAIL1>      ::= EOL | <OP> <2DTENSOR> <TAIL2> 
<TAIL2>      ::= EOL | <OP> <1DTENSOR> 
<2DTENSOR>   ::= "c(i, j)" | "c(j, i)" | "c(i, k)" | $\ldots$
<1DTENSOR>   ::= "b(i)" | "d(k)" 
\end{bnf}
\caption{An example generated template grammar, for the dimension list $[0,1,2,1]$, with a maximum of $3$ unique indices appearing in the candidate LLM solutions. 
The rules for each tensor index expression include all possible permutations of indices. 
 The symbol $\varepsilon$ denotes the empty string.
\label{bnf:bu}}
\end{figure}

\begin{algorithm}[t]
\caption{Bottom-Up Enumerator}
\label{algo:bu}
\begin{algorithmic}[1]
\Procedure{Enumerate}{$pCFG_{\template}, L$}
    \State $Q \gets \{(0,  pCFG_{\template}.S )\}$ \Comment{Priority queue initialized with start node }
    \While{$Q \neq \emptyset$}
        \State $(f, x) \gets Q.pop()$ \Comment{Remove template with minimal $f$}
        \If{$|tensors(x)| = |L|$} \Comment{If number of tensors in $x$ is the predicted number}
        \If{$x \notin \Sigma^*$}
        \State $x \gets \text{RemoveTail}(x)$  \Comment{Remove any tail nonterminal symbol}
        \EndIf
            \State $S \gets \textsc{Validate}(x)$ \Comment{Try to instantiate $x$}
            \If{$S \neq \bot$}
                \If{\textsc{Verify}$(x.\{S\})$} 
                 \State \Return $x.\{S\}$ \Comment{Return instantiated template}
                \EndIf
            \EndIf
        \EndIf
        \For {$x'$. s.t. $(x \xrightarrow{r} x' \wedge \,\, r \in pCFG_\template.R)$ } \Comment{Iterate over all possible expansions of $x$}
        \State $Q \gets Q \cup \{c(x') + g(x') + \Penalty(x'),  x'\}$
        \EndFor
    \EndWhile
    \State \Return Failure \Comment{Return Failure if no valid expression is found}
\EndProcedure
\end{algorithmic}
\end{algorithm}
The search algorithm maintains a queue of partial programs, as with the top-down search, which is initialized with the start symbol from the grammar. At each iteration, the program with the minimum cost function, as before, is popped from the queue, expanded, and all the resulting programs are added to the queue. 

The total cost function $f(x)$ for each partial expression $x$, is again defined as:
$f(x) = c(x) + g(x) + \Penalty(x),$
where $c(x)$ is calculated as before. In the bottom-up search, we use a simplified estimate of the cost to complete the program, $g(x)$, which is defined as:
$$g(x) = \sum_{i=k}^{|L|} m(L[i+1]),$$
where $k$ is the current number of tensors in $x$, $L$ is the predicted dimension list, and $m(d)$ is the minimal cost to add a tensor of dimension $d$. The minimal cost $m(d)$ is computed as follows, where Tensors($d$) is the list of tensors in the grammar of dimension $d$, and $\mathcal{P}[t]$ is the maximum probability of any production rule in the grammar which adds the tensor of dimension $d$:
    \begin{equation*}
    m(d) = -\log_2 \left( \max_{\tensor \in \text{Tensors}(d)} \mathcal{P}[t] \right).
    \label{eq:m_T}
    \end{equation*}

The penalty function is calculated as before, but with the criteria $\{b_1, b_2\} \in B$ defined as:
\begin{itemize}
\item $\Penalty_{b_1}(x) = 100$, where $b_1$ is violated if the tensor symbols in $x$ are not in alphabetical order by order of first appearance. 
\item $\Penalty_{b_2}(x) = \infty$, where $b_2$ is violated if $x$ contains at least as many tensors as predicted by the dimension list, and it uses fewer than half the operations available in the grammar, and $0$ otherwise.
\end{itemize}
The bottom-up search uses fewer penalty criteria than the top-down search because the construction of the grammar encapsulates a number of these criteria already (for instance, the tensors are enumerated by predicted dimension list order).

The main difference between the top-down and the bottom-up search is that the bottom-up grammar is generated in a way that, at each intermediate step, a complete program can be inferred from the partial program and checked against the specification. Every time an expression is dequeued from the queue, if the expression contains a tail nonterminal symbol, e.g., \texttt{\color{Plum}{TAIL1}}, \texttt{\color{Plum}{TAIL1}}; we can remove the tail nonterminal symbol to give a complete template (i.e., a template that contains no non-terminal symbols). We can then return this template to the template validator. If it fails validation, we will re-append the nonterminal symbol to the end of the expression and generate new expressions by expanding the non-terminal using all applicable production rules. The new expressions are added into the queue.

\section{Validation}
\label{sec:validation}
Once the synthesizer produces a complete template $\tau$, we wish to check whether it can generate a program $\target$ that satisfies the requirement $\forall \inputs. \src(\inputs) = \target(\inputs)$. 
Since checking this universally quantified formula is expensive, we first generate a set of tests in the form of input-output examples. 
This set of examples $\langle I, O \rangle$ is a list of input-output pairs where $I = \langle x_1 \mapsto v_1, ..., x_n \mapsto v_n \rangle$ is a map that binds the $\tau$ input arguments $\vec{x} = (x_1,...,x_ n)$ to concrete values $(v_1,...v_n)$ randomly generated and $O = (o_1,...,o_n)$ is the corresponding output produced when we execute $\src$ on the elements from $I$. 

The TACO templates generated during the synthesis phase contain symbolic placeholders for tensors and constants. To validate a candidate $\tau$ we build a set $S = (s_1,...,s_m)$ of substitutions $ s \mapsto x$ that map the symbolic symbols $s$ in  $tau$ to input arguments $x$. We iterate through all possible permutations of $S$, where tensor symbols are mapped to concrete tensor inputs, and constants are mapped to a set of constants $C = (c_1,...,c_m)$ containing the constant values present in the source code of $\src$. 
We can then generate a concrete program $\target = \tau.\{S\}$, and execute $\target$ on the input-output examples in $\langle I, O \rangle$. If any instantiated concrete program satisfies all the input-output examples, we return this to the next stage of verification.
We use $S$ to assign concrete values to TACO symbols and run $P_T$ to check its output.
When building $S$, we rule out invalid substitutions based on the type of arguments and TACO symbols. More specifically, we discard substitutions that try to bind tensor symbols with dimension $> 1$ to scalars and vice versa. 

If the validator succeeds, it returns the valid substitution $S$, if not it returns $\bot$ to indicate there was no valid substitution.

We explore all possible valid substitutions until we find a substitution $s^*$ that satisfies $\phi$. This validation process returns a tuple $\langle P_T, s^* \rangle$ that is given as input to the verification phase.

\begin{figure}[H]
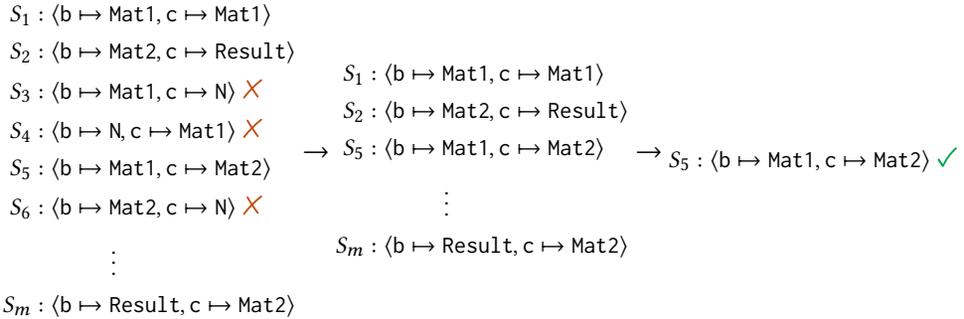
 
\begin{center}
    \begin{minipage}{0.25\linewidth}
        \small 
        \begin{align*}
            S_1 &: \langle \texttt{b}\mapsto \texttt{Mat1}, \texttt{c} \mapsto \texttt{Mat1} \rangle \\
            S_2 &: \langle \texttt{b}\mapsto \texttt{Mat2}, \texttt{c} \mapsto \texttt{Result} \rangle \\
            S_3 &: \langle \texttt{b}\mapsto \texttt{Mat1}, \texttt{c} \mapsto\texttt{N} \rangle \; \textcolor{Bittersweet}{\tikzxmark}\\
            S_4 &: \langle \texttt{b}\mapsto \texttt{N}, \texttt{c} \mapsto \texttt{Mat1} \rangle \; \textcolor{Bittersweet}{\tikzxmark} \\
            S_5 &: \langle \texttt{b}\mapsto \texttt{Mat1}, \texttt{c} \mapsto \texttt{Mat2} \rangle \\
            S_6 &: \langle \texttt{b}\mapsto \texttt{Mat2}, \texttt{c} \mapsto\texttt{N} \rangle \; \textcolor{Bittersweet}{\tikzxmark} \\
                &\;\;\;\;\;\;\;\;\;\;\;\; \vdots \\
            S_m &: \langle \texttt{b}\mapsto \texttt{Result}, \texttt{c} \mapsto \texttt{Mat2} \rangle
        \end{align*}
    \end{minipage}
     ${\rightarrow}$
    \begin{minipage}{0.25\linewidth}
        \small
        \begin{align*}
            S_1 &: \langle \texttt{b}\mapsto \texttt{Mat1}, \texttt{c} \mapsto \texttt{Mat1} \rangle \\
            S_2 &: \langle \texttt{b}\mapsto \texttt{Mat2}, \texttt{c} \mapsto \texttt{Result} \rangle \\
            S_5 &: \langle \texttt{b}\mapsto \texttt{Mat1}, \texttt{c} \mapsto \texttt{Mat2} \rangle \\
                &\;\;\;\;\;\;\;\;\;\;\;\; \vdots \\
            S_m &: \langle \texttt{b}\mapsto \texttt{Result}, \texttt{c} \mapsto \texttt{Mat2} \rangle
        \end{align*}
    \end{minipage}
     ${\rightarrow}$
    \begin{minipage}{0.25\linewidth}
        \small
        \begin{align*}
            S_5 &: \langle \texttt{b}\mapsto \texttt{Mat1}, \texttt{c} \mapsto \texttt{Mat2} \rangle\; \textcolor{Green}{\tikzcmark} 
        \end{align*}
    \end{minipage}
\caption{A set of possible substitutions for the TACO program  \texttt{a(i) = b(i,j) $*$ c(j)} and the inputs from the legacy program in Figure \ref{query:benchmark}. We discard invalid substitutions and try the valid ones until we find one that satisfies the specification.}
\label{fig:substitutions}
\end{center}
\end{figure}

\begin{example}
Figure \ref{fig:substitutions} shows a subset of the substitutions set $S$ given the program in Figure~\ref{query:benchmark} and the TACO candidate $P_T$ produced by the synthesizer, \texttt{a(i) = b(i,j) * c(j)}. Each substitution binds a symbol in the right-hand side of $P_T$, i.e., \texttt{b} and \texttt{c} to one of the inputs of \texttt{function}. The substitutions with a \textcolor{Bittersweet}{\tikzxmark} mark next to it are invalid, since they contain unsound bindings. For example, substitution $S_3$ binds \texttt{c}, a $1$-dimensional tensor to $N$, which is a scalar. Such substitutions are discarded and the valid ones are tested to run the program until we find one in which $P_T$ satisfies the specification. In this example, the correct substitution is $S_5$, which binds \texttt{b} and \texttt{c} to arguments \texttt{Mat1} and \texttt{Mat2} respectively.

\end{example}

\section{Verifier}
\label{sec:verifier}
We verify the correctness of a synthesized TACO program using bounded model checking. 
We compile both the original C and the TACO program to a common language within the MLIR compiler infrastructure~\cite{lattner2020mlir}. Given a TACO program $T$ and substitution $S$ returned by the validator, we create NumPy code based on the indexing expressions of $T$ and replace its variables for the concrete values specified in $s^*$. If the model checker fails to verify equivalence with the tuple $\langle T,s^* \rangle$, we return to the validation step and keep exploring different substitutions until we find one that satisfies the specification and passes verification.
We then use the JAX compiler \cite{jax2018github} to lower the NumPy code to MLIR.

From the MLIR files, we automatically generate C programs that create non-deterministic inputs, execute the original C and TACO code on copies of those inputs, and assert that the outputs are identical. We give this C program as input to CBMC \cite{Kroening2014}, a bounded model checker for C, that verifies said assertion holds for all possible inputs up to a certain bound.

Floating-point equivalence is both challenging to verify and, in many cases, undesirable. For instance, many compiler optimizations simply do not preserve floating-point optimizations, in order to achieve runtime speed-ups. For this reason, we extend CBMC to support rational datatypes, and verify equivalence using rational datatypes. 

\section{Evaluation}
To evaluate \sys and its various components, we compare its performance against several established techniques on a diverse suite of queries. The query set includes $10$ artificial examples and $67$ real-world problems ($61$ derived from codebases reported in the literature~\cite{c2tatco} and $6$ from the C++ based inference code of Llama~\cite{llama-bench}.)

\sys is implemented using an extended version of CBMC $6.3.1$ with cvc5 version $1.0.5$ as the underlying SMT solver. To generate initial candidate solutions, we use GPT-4 with the temperature set to $1.0$. A timeout of $60$ minutes is applied to each query. All experiments are conducted on a system equipped with an $11^{\text{th}}$ Gen Intel\textsuperscript{\textregistered} Core\texttrademark{} i5-1135G7 processor, $16$ GiB of RAM, and running Ubuntu\textsuperscript{\textregistered} $22.04.5$ LTS. Additional configuration details, including grammar refinements and penalty modifications, are provided in the subsequent sections.

We compare the following approaches: (\textbf{\sys$^{TD}$}) our approach, using the top-down \astar search described in Section~\ref{sec:top-down-weighted-astar};
    (\textbf{\sys$^{BU}$}) our approach, using the bottom-up search described in Section~\ref{sec:bottom-up-weighted-astar};
    (\textbf{C2TACO}) An enumerative synthesis tool for lifting C to TACO code~\cite{c2tatco}. We compare to C2TACO both with and without the domain-specific heuristics; 
    (\textbf{Tenspiler}) An enumerative synthesis tool based on the verified lifting framework~\cite{Qiu2024}; 
(\textbf{LLM only}) A baseline approach that employs a large language model (GPT-4) to directly generate candidate solutions without additional heuristic-driven refinement or search.
In addition, we perform ablation studies to evaluate the contribution of several components of \sys. Namely, the grammar refinement; the probabilities of the grammar; and the penalty functions.

We aim to answer the following research questions: 
\begin{description}
    \item[RQ1:] How does the performance of \sys compare to the state-of-the-art enumerative synthesis tools?
    \item[RQ2:] How does the choice of top-down or bottom-up A$^*$ search contribute to the performance of \sys?    
   \item[RQ3:] How much do the penalty functions contribute to the performance of \sys?
   \item[RQ4:] How much does the grammar refinement contribute to the performance of \sys?
   \item[RQ5:] How much do the probabilities on the grammar contribute to the performance of \sys?
\end{description}

\paragraph{\textbf{RQ1:} Comparison of performance of \sys to the state-of-the-art solvers: } 

\begin{figure}[h]
    \centering
    \includegraphics[width=0.8\linewidth]{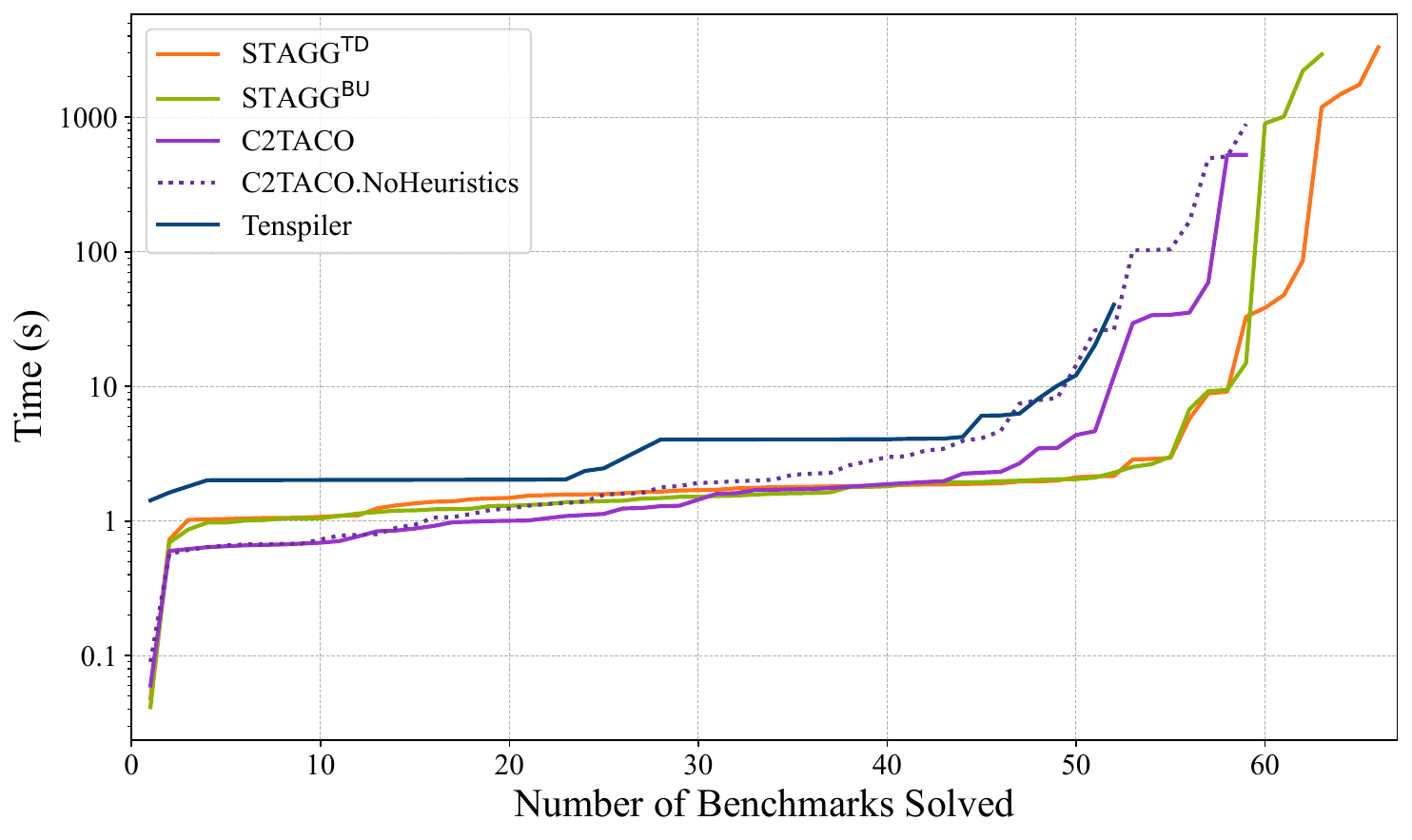}
    \caption{Cactus plot showing the number of benchmarks solved ($x$-axis) vs. time ($y$-axis, logarithmic) on the $67$ real-world benchmarks. Each line corresponds to a different synthesizer, and the point at which each line indicates how many benchmarks the synthesizer solved before the time.}
    \label{fig:cumulative}
\end{figure}

\begin{figure}
    \centering
    \includegraphics[width=0.8\linewidth]{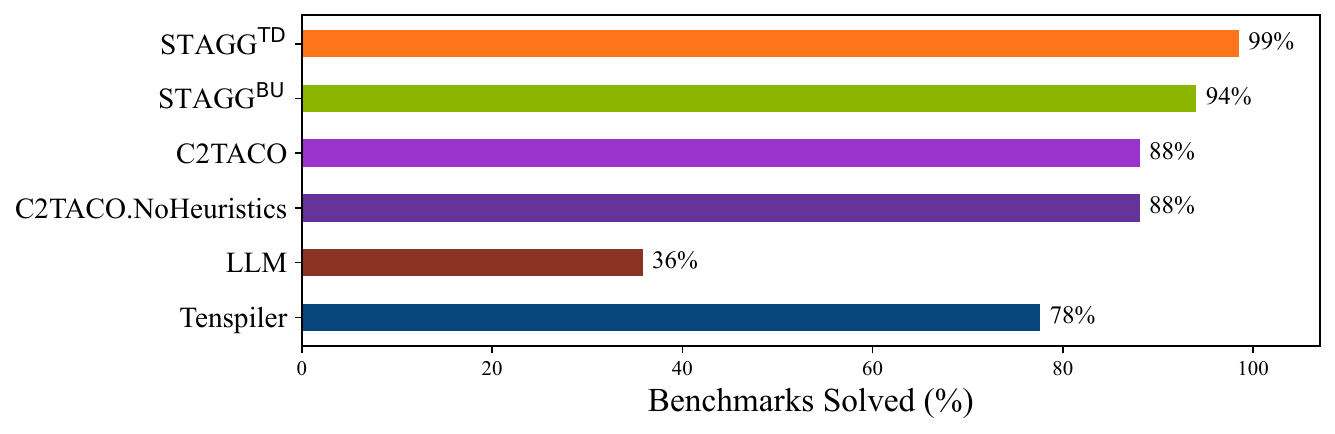}
    \caption{Success rates of different approaches on the set of $67$ real-world benchmarks.}
    \label{fig:barchart}
\end{figure}

Figure \ref{fig:cumulative} depicts the cumulative time each method takes over the $67$ real-world benchmarks, and Figure \ref{fig:barchart} shows the success rate of different techniques. We compare \sys to C2TACO on the full set of $77$ benchmarks in Table ~\ref{tab:total}. 
\systd solves $76$ benchmarks, compared to C2TACO which solves $67$. 
\systd solves all the benchmarks that C2TACO can solve, with an average solving time of $3.19s$, compared to C2TACO's $21.15s$. 
\sysbu solves $73$ benchmarks, and solves $66/67$ of the benchmarks that C2TACO solves, with an average solving time of $2.11s$ on the mutually solved benchmarks. C2TACO without the domain heuristics enabled is significantly slower.

We are only able to run Tenspiler on the $67$ real-world benchmarks, where it solves $52$. \systd solves all $52$ benchmarks that Tenspiler can solve, with an average time of $3.45s$ compared to Tenspiler's average time of $4.56s$. \sysbu solves only $50/52$ of the benchmarks that Tenspiler can solve, but with an average time of $2.03s$. This comprehensively answers RQ1: \sys outperforms the state-of-the-art solvers, both in terms of coverage and speed.

\begin{table}[]

\caption{
Comparison of benchmark-solving performance across different methods: The table reports the number of benchmarks solved (\#), average solving time (time in seconds), and attempts across various benchmarks. The benchmarks are categorized into real-world benchmarks (67 in total), real-world + artificial benchmarks (77 in total), benchmarks solved by C2TACO, and benchmarks solved by Tenspiler. \sys$^{TD}$ and \sys$^{BU}$ demonstrate superior solving capabilities, solving more benchmarks overall compared to C2TACO and Tenspiler, with \sys$^{BU}$ achieving the fastest solving times for benchmarks solvable by C2TACO and Tenspiler. The results highlight the efficacy of \sys over existing methods.}

\begin{tabular}{l|cc|ccc|cc|cc}
                    & \multicolumn{2}{c|}{\begin{tabular}[c]{@{}c@{}}Real-World \\ (67)\end{tabular}} & \multicolumn{3}{c|}{\begin{tabular}[c]{@{}c@{}}Real-World + Artificial \\ (77)\end{tabular}} & \multicolumn{2}{c|}{\begin{tabular}[c]{@{}c@{}}Solved \\ by C2TACO\end{tabular}} & \multicolumn{2}{c}{\begin{tabular}[c]{@{}c@{}}Solved \\ by Tenspiler\end{tabular}} \\ \hline
Methods             & \#                  & time                  & \#                  & time                  & attempts                  & \#                  & time                  & \#                  & time                  \\ \hline
STAGG$^{TD}$        & 66                  & 121.88                & 76                  & 106.13                & 44.55                     & 67                  & 3.19                  & 52                  & 3.45                  \\
STAGG$^{BU}$        & 63                  & 113.86                & 73                  & 98.81                 & 35.62                     & 66                  & 2.11                  & 50                  & 2.03                  \\
LLM                 & 24                  & 2.61                  & 34                  & 2.59                  & 1.62                      & 31                  & 2.57                  & 20                  & 2.72                  \\
C2TACO              & 59                  & 22.57                 & 67                  & 21.15                 & 18.45                     & 67                  & 21.15                 & 50                  & 23.69                 \\
C2TACO.NoHeuristics & 59                  & 43.08                 & 67                  & 49.41                 & 48.81                     & 67                  & 49.41                 & 50                  & 43.76                 \\
Tenspiler           & 52                  & 4.56                  &                     &                       &                           &                     &                       & 52                  & 4.56                  \\ \hline
\end{tabular}

\label{tab:total}
\end{table}

\begin{table}[]
\caption{
Impact of penalty rules on performance over 77 benchmarks (real-world + artificial). The table compares the number of benchmarks solved (\#), the percentage of benchmarks
solved (\%), and the average solving time (time in seconds) for various configurations of \sys. Removing penalty rules (e.g., Drop(A), Drop(B)) reduces the number of solved benchmarks and influences solving times. While \sys$^{TD}$ and \sys$^{BU}$ achieve high solving rates with the full penalty rules, dropping specific penalties often results in faster solving times but at the cost of reduced solving capability, as it failed solving complex benchmarks.}

\begin{tabular}{l|ccc}
                      & \multicolumn{2}{c}{Real-World + Artificial (77)} \\ \hline
Methods               & \#          & \%               & time                     \\ \hline
STAGG$^{TD}$          & 76          & 98.70\%          & 106.13                   \\
STAGG$^{TD}$.Drop(A)  & 71          & 92.21\%          & 7.21                     \\
STAGG$^{TD}$.Drop(a1) & 72          & 93.51\%          & 79.24                    \\
STAGG$^{TD}$.Drop(a2) & 75          & 97.40\%          & 91.66                    \\
STAGG$^{TD}$.Drop(a3) & 72          & 93.51\%          & 21.01                    \\
STAGG$^{TD}$.Drop(a4) & 75          & 97.40\%          & 90.58                    \\
STAGG$^{TD}$.Drop(a5) & 75          & 97.40\%          & 83.34                    \\
STAGG$^{BU}$          & 73          & 94.81\%          & 98.81                   \\
STAGG$^{BU}$.Drop(B)  & 70          & 90.91\%          & 68.18                    \\
STAGG$^{BU}$.Drop(b1) & 71          & 92.21\%          & 48.95                    \\
STAGG$^{BU}$.Drop(b2) & 70          & 90.91\%          & 68.75                    \\ \hline
\end{tabular}

\label{tab:penalty}
\end{table}

\begin{table}[]
\caption{
Performance comparison of different methods and grammar configurations over 77 benchmarks (real-world + artificial). The table shows the number of benchmarks solved (\#), the percentage of benchmarks solved (\%), the average solving time (time in seconds), and the number of synthesis attempts. \sys$^{TD}$ and \sys$^{BU}$ outperform C2TACO variants in solving more benchmarks. Variations of \sys demonstrate the impact of grammar refinement, where dropping penalty rules (Drop(A), Drop(B)) or using alternative configurations (e.g., EqualProbability, LLMGrammar) affect the solving capability, time, and attempts.}
\begin{tabular}{l|cccc}
                                               & \multicolumn{4}{c}{Real-World + Artificial (77)} \\ \hline
Methods                       & \#      & \%           & time        & attempts     \\ \hline
STAGG$^{TD}$                  & 76      & 98.70\%      & 106.13      & 44.55        \\
STAGG$^{TD}$.Drop(A)          & 71      & 92.21\%      & 7.21        & 13.65        \\
STAGG$^{TD}$.EqualProbability & 73      & 94.81\%      & 28.14       & 37.27        \\
STAGG$^{TD}$.LLMGrammar       & 52      & 67.53\%      & 3.77        & 5.25         \\
STAGG$^{TD}$.FullGrammar      & 69      & 89.61\%      & 91.15       & 874.29       \\
STAGG$^{BU}$                  & 73      & 94.81\%      & 98.81       & 35.62        \\
STAGG$^{BU}$.Drop(B)          & 70      & 90.91\%      & 68.18       & 10.07        \\
STAGG$^{BU}$.EqualProbability & 74      & 96.10\%      & 180.31      & 62.78        \\
STAGG$^{BU}$.LLMGrammar       & 52      & 67.53\%      & 2.74        & 2.60         \\
STAGG$^{BU}$.FullGrammar      & 68      & 88.31\%      & 96.57       & 259.35       \\
LLM                           & 34      & 44.16\%      & 2.59        & 1.62         \\
C2TACO                        & 67      & 87.01\%      & 21.15       & 18.45        \\
C2TACO.NoHeuristics           & 67      & 87.01\%      & 49.41       & 48.81        \\ \hline
\end{tabular}
\label{tab:different-grammar}
\end{table}

\paragraph{\textbf{RQ2:} Performance Comparison of Top-Down vs Bottom-Up Search} 
Our results show that, while \systd solves more benchmarks than \sysbu, \sysbu is faster on commonly solved benchmarks (for queries solved by both, \sysbu achieves a lower average solving time ($98.81$ seconds) compared to \systd ($108.83$ seconds)), and it enumerates fewer candidates. It has one big disadvantage though, which is that it can only expand expressions by appending to the previous expression, rather than by expanding nodes on the left-hand side of the AST. In particular, this means it cannot solve benchmarks that require expressions with more balanced Abstract Syntax Trees or benchmarks that contain parentheses.

\paragraph{\textbf{RQ3:} Contribution of the penalty functions}

Table~\ref{tab:penalty} shows the decline in performance for both \sys$^{TD}$ and \sys$^{BU}$ approaches when individual penalty rules are removed. As each penalty rule is dropped, the number of queries solved decreases, highlighting the importance of these rules in achieving high query-solving efficiency.

\begin{figure}
    \centering
    \includegraphics[width=0.8\linewidth]{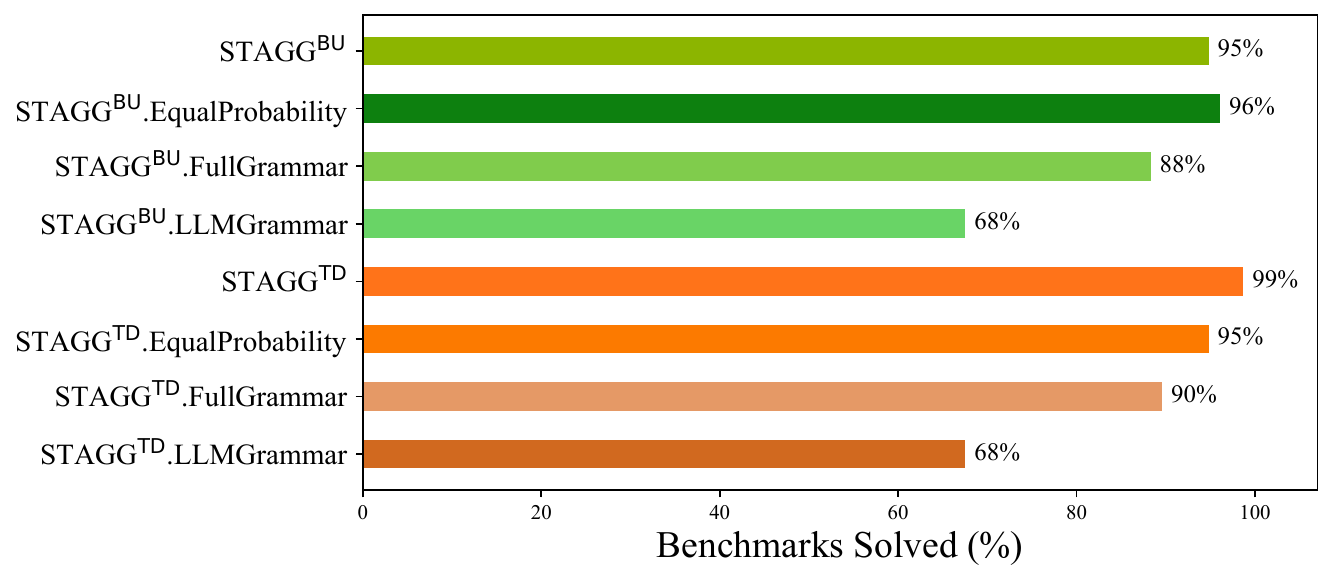}
    \caption{Impact of different grammar configurations in \sys on success rates across all $77$ benchmarks.}
    \label{fig:percentages-bar-grammar}
\end{figure}
\begin{figure}
    \centering
    \includegraphics[width=0.8\linewidth]{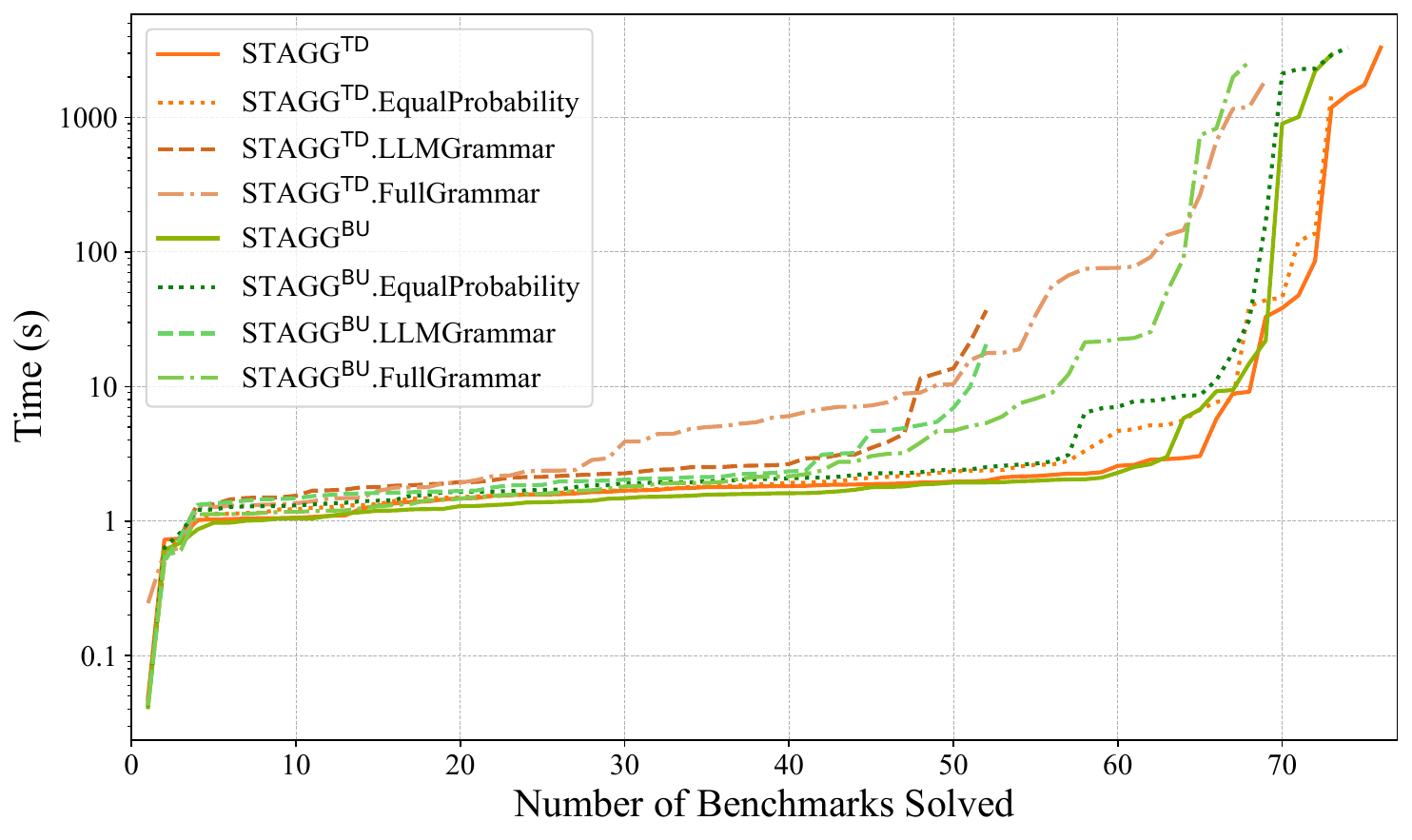}
    \caption{The performance of difference configurations of \sys on all $77$ benchmarks.}
    \label{fig:16}
\end{figure}

\paragraph{\textbf{RQ4} and \textbf{RQ5:} Contribution of grammar refinement and probabilities}
Figure~\ref{fig:percentages-bar-grammar},~\ref{fig:16} and Table~\ref{tab:different-grammar} show the performance of difference configurations of \sys: 
\texttt{EqualProbability} uses the refined grammar but replaces all probabilities in the generated pCFG with equal probabilities; \texttt{FullGrammar} uses the full TACO grammar in Figure~\ref{bnf:td} with equal probabilities; \texttt{LLMGrammar} uses the full TACO grammar in Figure~\ref{bnf:td} with probabilities learned from the LLM responses. All these configurations use the penalty functions. Thus, in order to compare the contribution of the grammar refinement, we can compare the performance of \texttt{LLMGrammar}, which uses learned probabilities but no refinement, to \sys. Dropping the refinement, here, results in us solving $31\%$ fewer benchmarks. If we compare \texttt{FullGrammar} to \texttt{EqualProbablity}, we can see that grammar refinement has less impact when learned probabilities are not used, but still results in a significant number of benchmarks being dropped. 

In order to compare the contribution of the probabilities, we can compare the performance of \texttt{EqualProbability} to \sys, where we note that using equal probabilities on the refined grammar results in an increase in the number of benchmarks solved for \systd, and an increase in solving speed for \sysbu, although neither is as impactful as the grammar refinement. In fact, the comparison between \texttt{FullGrammar} and \texttt{LLMGrammar} demonstrates that learned probabilities can have a negative impact if they are used in a grammar that is not general enough.
Thus, our answer to RQ4 and RQ5 is that grammar refinement in combination with probabilities has a bigger impact on performance than either component part, but the refinement alone is more powerful than the probabilities alone.

\section{Related Work}
\label{sec:related-work}

\subsection{LLM-guided Synthesis}
\label{sec:llm-guided-synthesis}

One of the strategies to aid the search process in the program synthesis tool is using a neural network to guide the search. Seminal work by Balog et al. ~\cite{Balog2016} synthesized array manipulation programs from I/O examples using a feedforward neural network (FNN) to build a probabilistic distribution over the target language. During the search, the synthesis algorithm expands partial programs based on the probabilities predicted by the FNN for the given I/O specification. Neural-guided synthesis has also been applied to solve string manipulation tasks by \cite{Odena2020, Shi2022Crossbeam}, inductive logic programming \cite{Shi2022Crossbeam}, dataframe \cite{Bavishi2019} and tensor \cite{shi2022tf,Nam2022} processing, and code transpilation \cite{Mariano2022} using distinct models. SketchAdapt \cite{Nye2019} uses a model to produce a program sketch as a starting point and completes said sketch through symbolic enumeration. Closer to our technique is Euphony \cite{Lee2018}, a system that learns a probabilistic high-order grammar model from examples and searches for programs using \astar. All those techniques require the model to be trained on domain-specific data, which can be costly and an impediment for such techniques to be expanded to other tasks.

Pre-trained Large Language Models (LLMs) have been used to guide enumerative synthesis in class program synthesis tasks:  
Li et al. ~\cite{LLM-SYGUS} developed two LLM-guided synthesis approaches to generate programs from logical formulas. The first method infers a probabilistic grammar using an LLM and enumerates said grammar using top-down \astar search; while the second queries the LLM to complete incorrect candidates produced by the enumeration phase. HYSYNTH~\cite{nadia} also derives a probabilistic grammar from LLM responses, but it uses bottom-up enumeration. 
LLMs are also used directly for verified lifting in LLMLift~\cite{Bhatia2024}, where lifting of programs is performed leveraging GPT-4 \cite{Achiam2023} to guess candidate solutions and loop invariants to prove equivalence, with feedback being given to the LLM to correct any mistakes. This approach is remarkably effective but relies on the LLM being able to fix its own mistakes based on the feedback, which may not be trivial in complex domains~\cite{selfrepair}.

\subsection{Tensor Code Lifting}
\label{sec:tensor-lifting} 

Driven by enormous advances in Machine Learning, there has been increasing interest in lifting tensor code to optimized targets. C2TACO \cite{c2tatco} is a synthesis tool that generates TACO code from I/O examples. It implements a bottom-up enumerative algorithm, and it uses code analysis to restrict the search space of programs. mlirSynth \cite{Brauckmann2023} has a similar approach, but it lifts tensor programs across different MLIR dialects. In both methods, correctness is asserted using only I/O testing while \sys performs bounded model checking to verify that the lifted programs are equivalent to their original counterpart. A different synthesis method was used in Tenspiler \cite{Qiu2024}, which employs symbolic synthesis to generate programs in six different tensor DSLs. Tenspiler builds verification conditions and loop invariants to prove that the lifted program is equivalent to the original one. Unlike \sys, which learns how to explore the search space in a fully automated way, all those techniques require hard-coded heuristics to make the search space tractable. 

Another approach to lift code is API matching, in which source code is replaced by optimized library routines to improve performance. Examples in the tensor domain include KernelFaRer \cite{DeCarvalho2021}, which focuses on general matrix multiplication (GEMM), and ATC \cite{Martinez2023}, which targets both GEMM and convolutions. SpEQ~\cite{SpEQ}  introduces a method of translating sparse linear algebra codes to optimized targets using equality saturation applied to LLVM IR. However, these approaches are often tailored to specific APIs and are not portable. \sys leverages the great learning capabilities of Large Language Models to infer the search space, which makes our technique extensible to different targets and to more unrestricted back-ends such as DSLs.

\section{Conclusions}
\label{sec:conclustion}
This paper presented \sys, a novel approach that combines LLMs and program synthesis to lift legacy tensor code to DSLs. We use a set of LLM responses to infer a probabilistic context-free grammar that drives an enumerative search over the space of possible solutions.
Our technique successfully lifts $99\%$ of a large suite of benchmarks with an average lifting time of $3.19$ seconds, outperforming existing state-of-the-art lifters in terms of coverage and synthesis time. Additionally, \sys is able to automatically learn a search space, and it does not rely on any pre-defined heuristics. Future work will focus on expanding our technique to application domains other than tensor computation.

\bibliographystyle{ACM-Reference-Format}
\bibliography{LLM-TACO}
\end{document}